\documentclass[useAMS,usenatbib]{mn2e}
\usepackage{graphicx}
\newcommand{\etal}{et al.~}

\newcommand{\Rvir}{R_{\rm vir}}
\newcommand{\Mvir}{M_{\rm vir}}

\newcommand{\Vvir}{V_{\rm vir}}

\def \kms {\ifmmode  \,\rm km\,s^{-1} \else $\,\rm km\,s^{-1}  $ \fi }
\def \kpc {\ifmmode  {\rm kpc}  \else ${\rm  kpc}$ \fi  }  
\def \hMsun {\ifmmode h^{-1}\,\rm M_{\odot} \else $h^{-1}\,\rm M_{\odot}$ \fi}
\def \hMpc {\ifmmode h^{-1}\,\rm Mpc \else $h^{-1}\,\rm Mpc$ \fi}
\def \hkpc {\ifmmode h^{-1}\,\rm kpc \else $h^{-1}\,\rm kpc$ \fi}
\def \hLsun {\ifmmode h^{-1}\,\rm L_{\odot} \else $h^{-1}\,\rm L_{\odot}$ \fi}

\newcommand{\mnras}{MNRAS}
\newcommand{\apj}{ApJ}

\newcommand{\apjs}{ApJS}
\newcommand{\aj}{AJ}
\newcommand{\aap}{A\&A}
\newcommand{\nat}{Nature}

\title[Galaxy groups and haloes in the SDSS-DR7]
{Galaxy groups and haloes in the SDSS-DR7}
\author[J. C. Mu\~noz-Cuartas, V. M\"{u}ller]
{Juan C. Mu\~noz-Cuartas $^{1}$; Volker M\"{u}ller $^{1}$\\
$^{1}$ Leibniz-Institut F\"{u}r Astrophysik Potsdam, An der Sternwarte 16, 14482
Potsdam, Germany.}
\begin{document}

\date{Accepted XXXX December XX. Received XXXX December XX; in original form
  2009 June 27}

\pagerange{\pageref{firstpage}--\pageref{lastpage}} \pubyear{2002}

\maketitle

\label{firstpage}

\begin{abstract}
  
In this work we introduce a new method to perform the identification
of groups of galaxies and present results of the identification of
galaxy groups in the Seventh Data Release of the Sloan Digital Sky
Survey (SDSS-DR7).  Our methodology follows an approach that resembles
the standard friends-of-friends (FoF) method. However, it uses
assumptions on the mass of the dark matter halo hosting a group of
galaxies to link galaxies in the group using a local linking
length. Our method does not assumes any ad-hoc parameter for the
identification of groups, nor a linking length or a density
threshold. This parameter-free nature of the method, and the
robustness of its results, are the most important points of our
work. We describe the data used for our study and give details of the
implementation of the method. We obtain galaxy groups and halo
catalogs for four volume limited samples whose properties are in good
agreement with previous works. They reproduces the expected stellar
mass functions and follow the expected stellar-halo mass relation. We
found that most of the stellar content in groups of galaxies comes
from objects with $M_r$ absolute magnitudes larger than -19, meaning
that it is important to resolve the low luminosity components of
groups of galaxies to acquire detailed information about their
properties.

\end{abstract}

\begin{keywords}
galaxies: haloes, groups -- cosmology: dark matter, observations
\end{keywords}


\section{Introduction}

It is well known that on small scales galaxies are distributed in an
inhomogeneous way. It is common to observe galaxies to be clustered,
forming groups and clusters of galaxies. Nowadays it is understood,
that the tendency of galaxies to cluster is a natural process
associated with their formation and evolution.

Galaxies are thought to form from the gas that cools in the potential
well of dark matter haloes. Posterior mergers between haloes induce
the growth of dark matter structures and influence the process of
galaxy evolution. Then, observing the spatial distribution of galaxies
allows for an indirect investigation of the spatial distribution of
the host dark matter haloes. Specifically, identification of galaxy
groups allows the identification of the dark matter structures that
host each group of galaxies.

Theoretically, dark matter haloes are associated with overdensities in
the dark matter density field and galaxies hosted in these haloes may
follow the local density enhancement. Then, from the observational
point of view, since dark matter can not be observed directly, the
overdensities in the mass density field have to be inferred from
enhancements in the local number density of galaxies. Unfortunately
there is not a general way to identify such enhancements, since
galaxies are a biased tracer of the mass density field and it is
difficult to establish a density threshold or a border that marks the
end of the distribution of galaxies that are associated to the same
dark matter halo. The situation becomes even more complex when one
considers the observational constraints on the data sets, like
incompleteness due to the non detection of faint galaxies, or
difficulties to resolve close pairs. Another difficulty comes from the
fact that observationally we can not determine the positions of
galaxies in real space. Because in galaxy surveys what we use to
determine the distance to a galaxy is its redshift, and it
encapsulates not only the effects of cosmic expansion but also
information about the dynamics of the local neighbourhood in which the
galaxy resides, the spatial distribution of the observational data
must be interpreted as in redshift space instead of real space. All
these inconvenients require the development of special techniques that
allow the identification of galaxy groups.

Now, because the distribution of galaxies can be considered as a point
process, the most straight forward way to identify groups of galaxies
in a survey is to use the FoF method. In this method the clusters are
identified using a percolation technique in which points are linked in
clusters if their mutual and transitive distances are smaller than $b$
times the mean interparticle distance. In numerical simulations,
where the particles of the point process represent mass elements with
a well known mass, one can choose the value of $b$ in order to select
regions that are bounded by some given overdensity threshold. In
observations, since the point process represents galaxies, and for
instance there is no simple way to assign masses to each galaxy, the
selection of the value of the linking length has to be done on a
empirical basis, and only after tests one can choose a value that
gives confident results (Berlind \etal 2006). Furthermore, in redshift
space, due to the break in the spatial symmetry introduced by the
redshift space distortions, one has to split the search in two
orthogonal directions and then use two different linking lengths whose
values have to be tuned upon the performance of tests.

In no way the identification of groups of galaxies as described above
is a warranty of genuine group selection, and in the basic picture, it
is not possible to obtain further information about their dark matter
haloes.

Previously, exploiting the wealth of data provided by the already
existing surveys, many works have focused on the identification of
groups of galaxies in galaxy redshift surveys. For instance, Berlind
\etal (2006) have performed the identification of groups of galaxies
in the third data release of the Sloan Digital Sky Survey. Besides the
groups of galaxies and their properties, they have shown in their work
a detailed study on the effects of the selection of the linking
length, finding an optimal value that allows them to study halo
occupation statistics. Later Crook \etal (2007), using a percolation
method, presented the identification of groups of galaxies in the Two
Micron All Sky Redshift Survey (2MASS). Their samples have been
designed to maximize the number of rich groups and those required to
be identified above some given overdensity threshold. They also
present a match between the most massive groups in their catalog with
previously well known groups and clusters of galaxies.

Merch\'an \& Zandivarez (2005) have proposed a standard implementation
of the FoF method in flux limited samples. In their implementation
they avoid artificial merging of small groups and perform an improved
determination of the group center for rich groups. They implemented
the method on the third data release of the SDSS. Zapata \etal (2009)
and Zandivarez \& Mart\'inez (2011) also have used this prescription
to perform the identification of groups in different releases of the
SDSS. Particularly, Zandivarez \& Mart\'inez (2011) have used this
prescription to identify groups of galaxies with at least four
members. They have used linking lengths that correspond to
overdensities comparable to those used to define dark matter
overdensities in standard $\Lambda$CDM cosmology and computed the
virial halo mass from an estimated virial radius and the velocity
dispersion of member galaxies.

Koester \etal (2007) have used the MaxBCG method to identify clusters
of galaxies in the SDSS. The method, based on the likelihood
associated to a galaxy to be a BGC and a likelihood associated to the
spatial, morphological and photometric properties of the galaxy, uses
a percolation algorithm to identify groups. Particularly, they show a
high purity of the clusters they identify with this method. Geach
\etal (2011) have used a technique based on a Delaunay tessellation on
sets of galaxies distributed by colour. With this method, they
identify photometrically selected clusters of galaxies out to a
redshift $z \sim 0.6$. Although this method allows the identification
of potential clusters at high redshift, it is difficult to actually
confirm the physical relation among cluster members. Lee \etal (2004),
Nichol (2004), Wen \etal (2009), Tago \etal (2008) and Tago \etal
(2010), have also shown the identification of groups of galaxies in
different data releases of the Sloan Digital Sky Survey.

Of particular interest for our work has been the group catalog
presented in Yang \etal (2007) (hereafter YHC). In YHC, on top of a
standard FoF group identification, they perform a halo based
identification of groups of galaxies hosted in the same dark matter
halo. Several different studies have shown the good performance of
such a group finder. Their halo catalog, which is based on the data of
the Fourth Data Release of the SDSS, has been used in different works
to study the conditional luminosity function (Yang \etal 2005a, Yang
\etal 2005b, Wang \etal 2008), environmental properties of galaxies
(Wang \etal 2011, Wetzel \etal 2011, Weinmann \etal 2011 and Wang
\etal 2008), the distribution of dark matter on large scales
(Mu\~noz-Cuartas \etal 2011), among others. Recently Tinker \etal
(2011) have used the same method to identify the groups of galaxies in
the SDSS-DR7 to study the galaxy-halo connection.

In this work we revisit the problem of the identification of groups of
galaxies in galaxy redshift surveys, and the identification of dark
matter haloes from the groups of galaxies they host. Our method is
inspired by the method of Yang \etal (2007), and follow the ideas of
hierarchical growth of structures for the assembly of dark matter
haloes as well as the ideas of the standard FoF method.

We assume that each galaxy has an associated dark matter halo, with a
mass that depends on the luminosity or the stellar mass of the galaxy
it hosts. Then we use the estimated halo mass to compute its
properties, and make the search of neighbouring galaxies in an
ellipsoidal region, with axes determined by the virial radius of the
halo and its maximum circular velocity. In this way, we merge groups
that intercept the ellipsoid of a given halo, in a similar way as the
FoF method, with the difference that the linking length is local and
completely dependent on the properties of the dark matter halo that is
being the current center of search.

Our procedure provides two major improvements on previous percolation
methods. First, the linking lengths for search of neighbouring
galaxies is local, and it depends only on the properties of the halo
that is the center of search. And second, and more important, no
assumption about the value of the initial linking length or any other
parameter has to be made for the identification of galaxy groups and
haloes. Therefore our method is parameter free. Furthermore, like the
method of Yang \etal (2007), the identification of groups of galaxies
leads directly to the identification of dark matter haloes in the
survey with reliable mass estimates. On the other hand, our method
differs from that presented in Yang \etal (2007) in several
aspects. First, we do not need to make an initial FoF procedure in
redshift space to start the iterations of the groups. This avoids
possible contamination by the choice of the initial linking
lengths. Second, differently from the assumption made in Yang \etal
(2007), we do not need to assume an initial value for the
mass-to-light relation of groups. And third, our method uses a two
dimensional spheroid for the search of group members, much in the way
as the FoF algorithm with two orthogonal adaptive linking
lengths. This is different from the implementation in Yang \etal
(2007), where they use a fixed FoF linking length for the transverse
search and a probabilistic approach for the redshift distribution of
galaxies, that at the end requires the use of another free parameter
to fix the density contrast defining the membership of galaxies in
groups.

In what follows, in Section \ref{sec:method} we introduce the method
for identification of groups of galaxies. Then in section
\ref{sec:data} we present the data set used to implement the method
presented in the work. Then in section \ref{sec:results} we present
the results of the implementation of our method on the data of the
Seventh Data Release of the SDSS. Finally we summarise and discuss our
results.


\section{Group Identification Method} 
\label{sec:method}

Our group finder is based on the idea of haloes, and is inspired in
the philosophy of a standard FoF group finder as well as from the
group finding method proposed by Yang \etal (2007).

In the standard FoF method, one usually chooses a fixed linking length
that is a fraction $b$ of the mean inter-particle distance. Particles
that are closer than this linking length are labelled as members of the
same cluster, where the membership is transitive through all particles
in the data set, this means that through the particle distribution,
{\em friends of my friends are also my friends}. In N-body
simulations one can show that using a linking length $b=0.2$ times the
mean interparticle separation, the overdensity of the structures
identified is around $170$ times the mean density of the universe, a
number that is in close agreement with the expectation from spherical
collapse model for virialized structures. Also, because of the
isotropy of the particle distribution, only one linking length is
required for the search in the 3D space.

\begin{figure}
  \includegraphics[width=7.8cm,angle=270]{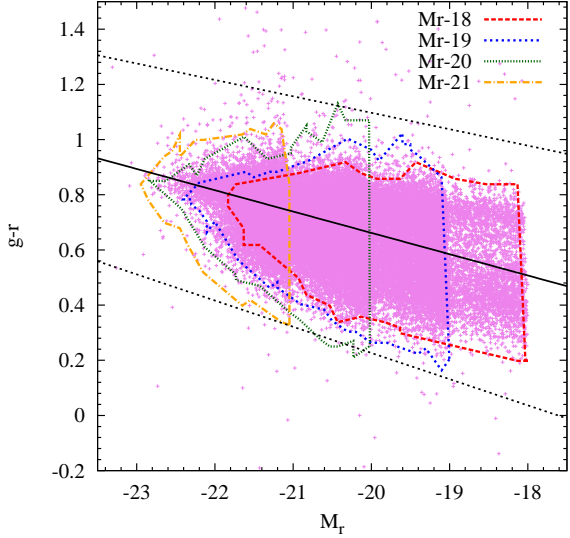}
  \caption{Colour magnitude diagram for the galaxies in the four
    volume limited samples. The straight lines represent the mean and
    $3\sigma$ levels used to adjust the colour used in the estimation
    of the stellar mass $M_*$ for those colour outliers. The contours
    indicate the regions in the diagram delimited by each galaxy
    sample.}
  \label{fig:Galaxycolors}
\end{figure}

In the case of observational data points the search has to be split in
a two dimensional problem because the redshift space distortions
breaks the spatial symmetry of the distribution of points
(galaxies). Then it is required to make the search in two orthogonal
directions using two different linking lengths. In this case the
selection of both linking lengths becomes arbitrary. First, because
now the data points do not represent mass elements, but galaxies, for
which we do not know a-priory the amount of mass they represent,
i.e. one can not use the same arguments as in simulations to determine
the value of the linking lengths. Second, the amount of distortion
introduced by redshift space effects is unknown, and it depends on
different factors like the host halo mass, the mass of the satellite
galaxy, and their positions relative to the observer. In this case, to
obtain reliable results on the identification of groups, one has to
look for the set of parameters that gives the best results according
to the expected properties of the clusters (Bell \etal 2006, Tago
\etal 2005, Tago \etal 2010).

We aim to perform the search of groups of galaxies residing in the
same dark matter halo. Our approach uses a local linking length
criteria, that depends only on the properties of the assumed host dark
matter halo associated with every galaxy group. Now we describe our
method step by step.\\


{\it Initialization:}

Before start, we have to prepare a set of quantities for each galaxy
in the galaxy catalog. We compute the galaxy's stellar mass,
luminosity in the $r$ band, comoving Cartesian coordinates and
estimate the halo mass and radius.

We first compute for each galaxy its stellar mass $M_*$. Following
Yang \etal (2007) we do so through

\begin{equation}
  \log(M_*) = -0.306 + 1.097(g-r)^{0} - \epsilon + \log(L_r)
  \label{eq:MstellarGal}
\end{equation}

\noindent
where we have used the relations from Bell \etal (2003) to estimate
the stellar mass from the mass to light ratios. $(g-r)^{0}$ is the
colour of the galaxy corrected to $z=0$, $\epsilon$ is a parameter
that depends on the initial mass function, that in our case is
$\epsilon=0.15$ (Bell \etal 2003), and $L_r$ is the luminosity of the
galaxy in the $r$ band, that has been computed as

\begin{equation}
  L_r = 10^{0.4(M_{\odot,r} - M_r)}
  \label{eq:Lr}
\end{equation}

\noindent
with $M_{\odot,r}=4.76$ (Blanton \& Roweis 2007). As discussed in Yang
\etal (2007), some galaxies in the catalog are outliers of the colour
magnitude diagram. Using these values for the colours in
eq. \ref{eq:MstellarGal} will produce unreliable stellar masses for
these galaxies. We do not know the reason of the exceptional behaviour
of these galaxies, and therefore it is not clear how to assign a
colour to them. For this reason we assume the simplest approach. For
galaxies that are $3\sigma$ values off from the mean value of
$(g-r)^{0}$ in the colour magnitude diagram, we estimate their stellar
masses using the mean value for the colours of the galaxies with the
same $L_r$ luminosity. Figure \ref{fig:Galaxycolors} shows the colour
magnitude diagram for our four volume limited samples as well as the
mean and $3\sigma$ levels. As it can be seen, only a small fraction of
the galaxies in the sample need their stellar masses to be corrected
for colour.

For every galaxy we compute also their comoving Cartesian coordinates
from

\begin{eqnarray}
x_i &=& r_i\cos(\delta_i)\cos(\alpha_i), \nonumber\\
y_i &=& r_i\cos(\delta_i)\sin(\alpha_i), \\
z_i &=& r_i\sin(\delta_i), \nonumber
\end{eqnarray}

\noindent
where $\alpha_i$ and $\delta_i$ are the right ascension and
declination of each galaxy in the catalog and $r_i$ is the comoving
distance of the respective galaxy, given by

\begin{equation}
  r_i = c \int_0^{z_i}\frac{dz}{H_0\sqrt{\Omega_m(1+z)^3 + \Omega_{\Lambda}}},
\end{equation}

\noindent
with $c$ being the speed of light, $\Omega_m=0.258$ is the mass
density parameter and $H_0=72~\kms\rm{Mpc}^{-1}$ is the Hubble
constant at present time.

Then, we assume that every galaxy resides in a dark matter halo of a
given mass. We assign mass to the dark matter halo of each galaxy
relating the stellar mass (or the $L_r$ luminosity) of the galaxy with
the mass of dark matter haloes sampled from a theoretical mass
function (Warren \etal 2006, Sheth \& Tormen 2002) between mass limits
$M_{min}$ and $M_{max}$. Later we will describe how we define these
limiting values.

Once each halo has a mass, we compute their radius assuming they
satisfy virialization criteria. Then each halo has a radius $\Rvir$
given by

\begin{equation}
  \Rvir=\left(\frac{3\Mvir}{4\pi
    \Delta_\mathrm{vir}\rho_\mathrm{crit}} \right)^{1/3}.
  \label{eq:Rvir}
\end{equation}

We have assumed that the mass assigned to the halo coincides with its
virial mass, and use the mean overdensity of haloes in the spherical
collapse $\Delta_{\rm vir}$ relative to the critical density of the
Universe, $\rho_\mathrm{crit}$, evaluated at $z=0.1$ (Bryan \& Norman
1998). After this, we sort the halo catalog in decreasing order of
mass to start iterations from the most massive haloes to the low mass
ones. In this initial situation each halo hosts one galaxy. We set the
position of the halo to coincide with the position of the galaxy it
hosts.\\

{\it Iteration:}\\

{\it Step one:} Once there is an initial halo catalog (haloes with
positions, radius and masses) we can start linking haloes that are
close to each other. Starting from the most massive halo ($S_h$) we
search for the haloes ($S_i$) that are contained in a sphere of radius
$R_{zs}$ centred at the position of halo $S_h$. The radius of the
sphere $R_{zs}$ is given by

\begin{equation}
  R_{zs} = \frac{V_{\rm max}}{100} ~ [\hMpc]
  \label{eq:Rzs}
\end{equation}

\noindent
where $V_{\rm max}$ is in units of \kms and represents the median
maximum circular velocity of the halo $S_h$ with mass $\Mvir$ that is
center of search. $V_{\rm max}$ is approximated by

\begin{equation}
  V_{\rm max} = 0.0325\left(\frac{\Mvir}{M_{\odot}}\right)^{0.31} [\kms]
\end{equation}

\noindent
as computed from high resolution cosmological simulations
(Mu\~noz-Cuartas \etal in prep.). Instead, we could have used the halo
velocity dispersion $\sigma_v$, which would have been more appropriate
from the theoretical point of view. However we tested both
approximations and the differences are small since both quantities are
comparable. On the other hand, using $V_{\rm max}$ produces a region
of search $R_{zs}$ that is slightly larger than the one using
$\sigma_v$. We prefer a larger $R_{zs}$ to maximize the number of
members per group with the hope that using a large $R_{zs}$ the
effects of redshift space distortion are treated more carefully.
Finally, tests against mock catalogs have shown that this choice
produces the results with the highest purity and completeness,
compared to $\sigma_v$ or $\Vvir$.\\

\begin{figure*}
  \includegraphics[width=7.8cm,angle=270]{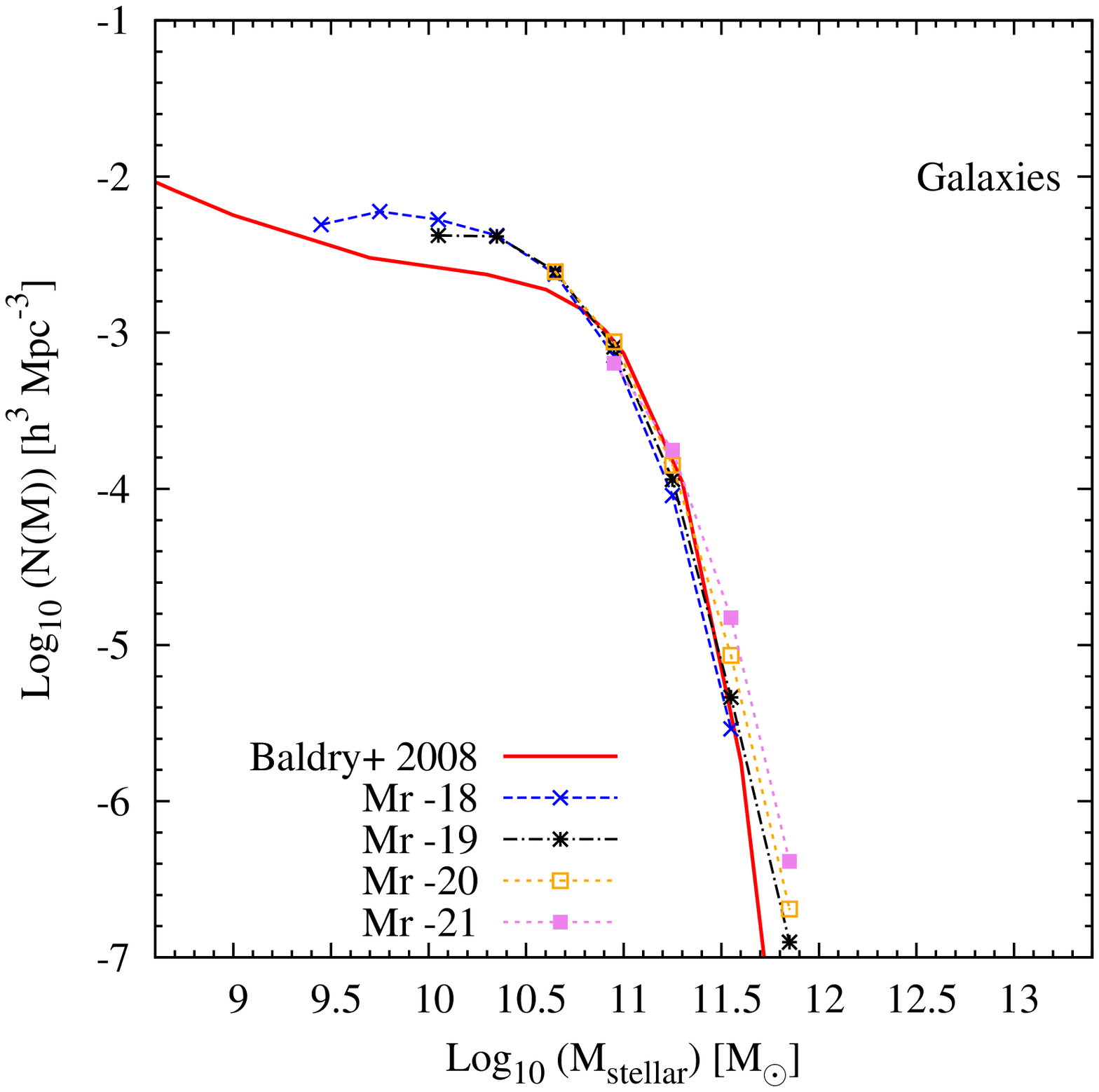}
  \includegraphics[width=7.8cm,angle=270]{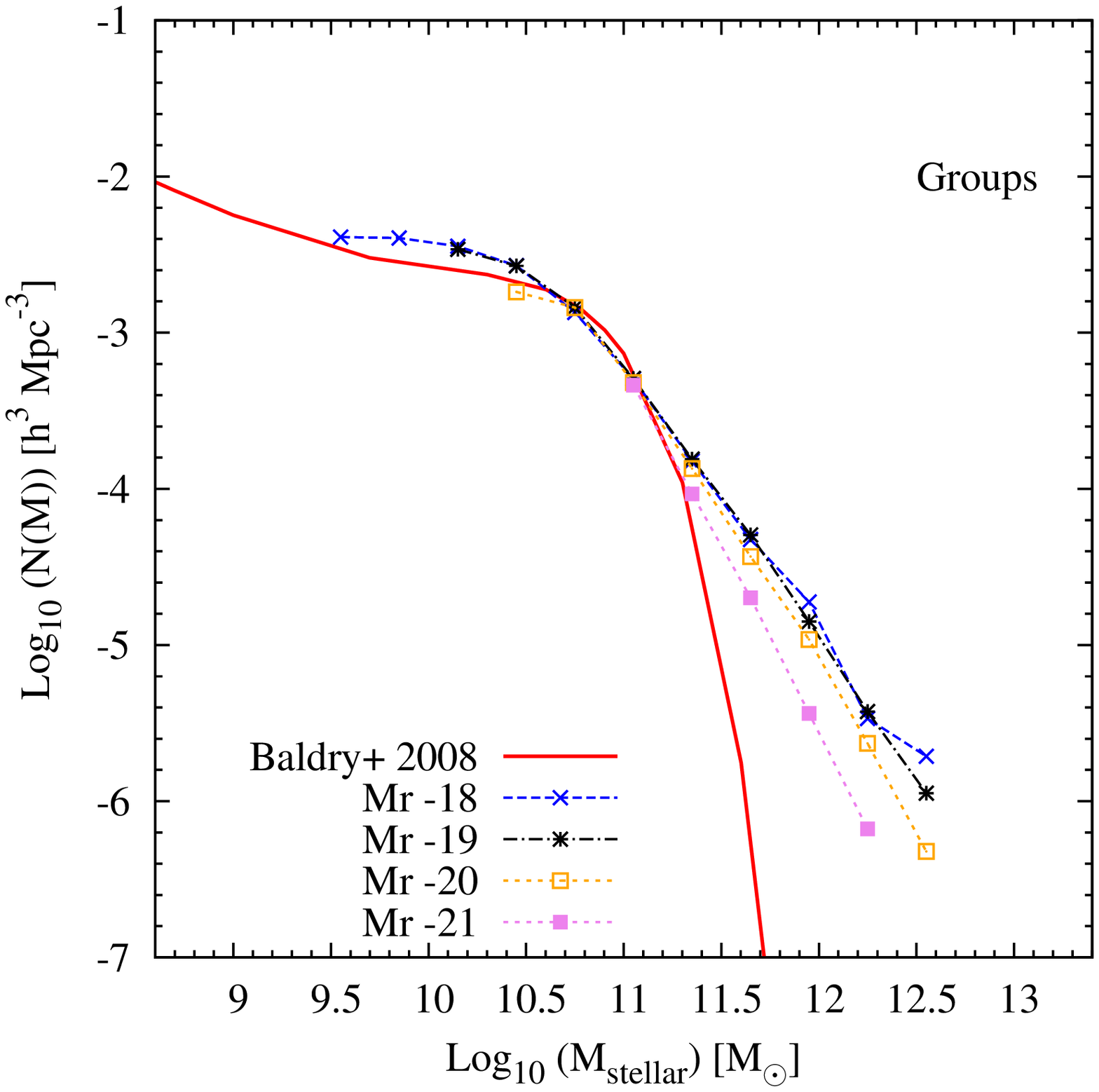}
  \caption{Stellar mass functions for galaxies and groups for our four
    volume limited samples. The solid line shows the stellar mass
    function of Baldry \etal (2008).}
  \label{fig:StellasMassFunc}
\end{figure*}

For the set of haloes $S_i$ that are inside the sphere of search
$R_{zs}$ we perform a set of operations:

\begin{enumerate}
\item We evaluate their positions relative to the position of the halo
  $S_h$

\item Rotate the coordinates of the haloes in $S_i$ to a system of
  coordinates such that the line of sight coincides with the $z$-axis.

\item Search for the subset of haloes $S_m$ in $S_i$ that are
  contained in the spheroid defined by

\end{enumerate}

\begin{equation}
  f(\Rvir,R_{zs}) = \frac{x^2}{\Rvir^2} + \frac{y^2}{\Rvir^2} + \frac{z^2}{{R_{zs}}^2} .
\end{equation}

\noindent
Here $\Rvir$ is the virial radius of the halo $S_h$.

Now the meaning of the quantity $R_{zs}$ becomes clear. $V_{\rm max}$
represents the maximum circular velocity of a particle in the
potential of the host halo $S_h$, then it sets a limit for the
velocity of a galaxy moving inside the halo. If a galaxy moves with a
peculiar velocity $V_{\rm max}$ along the line of sight, then its
observed velocity will be $V_{obs} = V_{\rm max} + V_H$ where $V_H$ is
the Hubble flow at the position of the halo $S_h$. Then, if we
approximate the distance to the galaxy as $d=(V_{\rm max} + V_H)/H_0$
it becomes clear that $R_{zs}$ will represent the maximum distance
along the line of sight from the center of the halo $S_h$ at which we
can find a galaxy bound to its potential well.

Note that the use of an ellipsoidal window function accounts for the
projection effects of the peculiar velocities, which are maximum along
the line of sight, but reduce gradually out of this
line. Particularly, note that galaxies (and haloes) with no projected
velocities along the line of sight, but at a distance $\Rvir$ from the
center of the halo (galaxies moving perpendicular to the line of
sight) will be included as members of the group. This process tries to
recover the spatial symmetry of the point process that is broken by
the redshift space distortions. It also avoids the ad-hoc splitting of
the search in to two orthogonal directions with two different linking
lengths with no physical relation between them.

{\it Step two:} All haloes in the subset $S_m$ that fall inside the
spheroid delimited by $f(\Rvir,R_{zs})$ are merged with the halo
$S_h$. In this process, all galaxies inside the small haloes in $S_m$
will be incorporated in the halo $S_h$. The position of the resulting
halo will be relocated to the position of the galaxy with the largest
stellar mass. The final mass of the halo right after merging will be
the sum of the masses of the haloes that merged.

Total stellar masses in groups are then computed as the sum of the
stellar mass of the individual galaxies $M_{*,i}$ weighted by their
completeness $c_i$

\begin{equation}
  M_{stellar} = \sum_i \frac{M_{*,i}}{c_i}
  \label{eq:Mstellar}
\end{equation}

\noindent
Similarly, characteristic luminosities in the $r$ band, $L_{ch}$, of
each group are computed from the luminosity of their galaxy members as

\begin{equation}
  L_{ch,g} = \sum_i \frac{L_{r,i}}{c_i}
\end{equation}

\noindent
As discussed in Yang \etal (2007), weighting the stellar mass and
luminosity with the completeness accounts for the missing objects in
the same region of the survey.

Haloes that are already merged in larger haloes are removed from the
halo catalog for the next iteration. Then, during the procedure, the
halo population changes but not the galaxy population.\\

{\it Step three:} After applying this procedure to all haloes $S_h$,
in decreasing order of mass, and considering that the population of
haloes will change each time small haloes merge with the big ones, we
end up with a new distribution of galaxies in haloes and a new halo
catalog. We then update the halo list and sort the halo catalog in
decreasing order of mass.\\

{\it Step four:} Once the halo catalog is sorted, we assign masses to
haloes. We assume a one-to-one correspondence between halo mass and
group stellar mass or group characteristic luminosity $L_{ch}$. Using
the estimated comoving volume of the sample of galaxies, we compute a
minimum $M_{\rm min}$ and a maximum mass $M_{\rm lim}$ necessary to
host all haloes in the halo catalog. The mass limits are obtained from
the mass function solving the equations $N(>M_{\rm lim})=1$ and
$N(>M_{\rm min})=N_{h}$, where $N(>M)$ is the cumulative mass function
representing the number of haloes with mass larger than $M$ and
$N_{h}$ is the number of haloes in the halo catalog at a given
iteration.

Because during the first iterations (for the initialization step, and
a few initial iterations) we associate mostly individual galaxies with
individual haloes, the use of the limit halo mass $M_{\rm lim}$ will
associate unreliable massive haloes to individual galaxies, for
example, galaxies with a stellar mass of $\sim10^{12}$ \hMsun could be
associated with a halo with a mass of $>10^{14}$ \hMsun. This high
mass will lead to large linking lengths $\Rvir$ and $R_{zs}$ to these
haloes, i.e, the massive haloes would grow very rapidly and would
hinder the identification of smaller haloes. This would have a
negative effects on the final distribution of groups and haloes. To
avoid it we fix, for each iteration, an ad-hoc maximum halo mass
$M_{\rm max} < M_{\rm lim}$ and sample randomly $N_{h}$ halo masses
from the mass function in the range of masses $[M_{\rm min},M_{\rm
    max}]$. Then we iteratively shift the value of $M_{\rm max}$ by
some amount $dM$ until it reaches the maximum allowed halo mass in the
volume, $M_{\rm lim}$. In this way we do a physically reasonable mass
assignment to massive groups in each iteration, and control the growth
of the linking lengths of haloes.

Then, we generate $N_{h}$ halo masses sampling the mass function in
the interval $[M_{\rm min}$, $M_{\rm max}]$. Then, halo masses are
assigned, in each iteration, relating them to the galaxy groups
according to the group stellar mass $M_{stellar}$ or the group
characteristic luminosity $L_{ch}$, in a way that the groups with the
largest $M_{stellar}$ or $L_{ch}$ will be associated to the most
massive halo mass.

{\it Iterate all steps:} We iterate the previous steps (one to four)
for a given pair of values of $M_{min}$ and $M_{max}$ starting with
$M_{max}=10^{12}$\hMsun. Once this iteration converges to a fixed
number of haloes, we increase the value of $M_{max}$ by an amount
$dM=\Delta\log_{10}(M)=0.5$. Since the number of haloes in the
population have changed, we recompute the value of $M_{min}$, and
repeat the process from the first step with these new values of
$M_{min}$ and $M_{max}$. This process has to be repeated until
$M_{max} = M_{lim}$.

We have found that a reasonable value for $M_{max}$ to start the
iteration is $10^{12}$\hMsun, which approximately corresponds to the
mass of the halo of massive galaxies at $z=0$. We have tested
different values of the starting $M_{max}$ and $dM$ and found no
differences in the final population of groups. In conclusion our
results are weakly dependent on the choice of $M_{max}$ and $dM$. The
independence of the final results on $M_{max}$ and $dM$ is due to the
double-iteration process that controls the evolution of the population
of haloes.

Note that our method does not assumes any ad-hoc parameter for the
identification of groups, nor a linking length or a density
threshold. The parameter-free nature of the method, and the robustness
of its results are one of the most important points of our work.



\section{The Data  Sample} 
\label{sec:data}

Our galaxy groups are identified from the SDSS Seventh Data Release
(Abazajian \etal 2009). We use data for the SDSS-DR7 publicly
available from the VAGC (Blanton \etal 2005). The data release from
the VAGC has improvements on the photometric reductions, calibration
as well as deals with the problems associated with multiple
observations of the same object. To produce our fiducial galaxy
catalog from the VAGC catalog we first select objects that are
targeted simultaneously as ``GALAXY'' and have been targeted in the
spectroscopic survey. Furthermore we request the object to belong to
the main galaxy sample (``VAGC\_SELECT=7'') and to be a well resolved
spectral target (``RESOLVE\_STATUS\&256''). Our initial galaxy sample
extends from $z=0.002$ to $z=0.2$. For all galaxies in this sample we
compute the $K$-corrected absolute magnitudes using the KCORRECT code
(V4.2, Blanton \etal 2007). Absolute magnitudes are computed as shown
in Yang \etal (2007), from

\begin{equation}
   M_x - 5\log{h} = m_x - DM(z) - K(z) - A_x(z - z_{n})
\end{equation}

\noindent
where the subscript ``$x$'' stands for the different magnitude band
$x=(u, g, r, i, z)$, $M_x$ is the absolute magnitude of the galaxy,
$m_x$ is the apparent magnitude of the galaxy as given in the catalog,
$DM(z)$ is the distance modulus, $K(z)$ is the $K$-correction term and
$z_{n}=0.1$ is the reference redshift. The coefficients $A_x$ quantify
the correction for evolution, and are taken from Blanton \etal (2003)
to be $A_x=(-4.22, -2.04, -1.62, -1.61, -0.76)$ for the five colours.

Following Tago \etal (2010), we generate four different volume limited
samples, taken from the previously described galaxy sample, each with
magnitude cut $M_r=-18$, $-19$, $-20$ and $-21$. The redshift limits
are $z_{min}=0.002$ and $z_{max}=0.047$, 0.074, 0.115 and 0.175
respectively. We have imposed a limit at $M_r=-23.5$ as the maximum
absolute magnitude of a galaxy in the sample. No more than ten
galaxies are discarded at this step.  Table \ref{tab:tab1} summarizes
the properties of the four volume limited samples used in this
work. The total comoving volume occupied by each galaxy sample is
estimated using a Delaunay tessellation, which allow us to estimate
the volume of each galaxy, and then the total volume as the sum of the
volumes of all galaxies in the sample\footnote{We have used the
  publicly available code Qhull (Bradford \etal 1996) to compute the
  Delaunay tessellation.}.

\begin{table}
\begin{center}
\begin{tabular}{|c|c|c|c|c|}\hline \hline

Name   & $z_{min}$ & $z_{max}$ & $N_{gals}$ & $L_{eq}$ \hMpc   \\\hline
Mr-18  & 0.002     & 0.047     & 50986      & 130   \\
Mr-19  & 0.002     & 0.074     & 108546     & 200   \\
Mr-20  & 0.002     & 0.115     & 155890     & 310   \\
Mr-21  & 0.002     & 0.175     & 97064      & 465   \\
All    & 0.002     & 0.2       & 412486     &       \\

\hline
\hline
\end{tabular}
\end{center}
\caption{Summary of the properties of the volume limited samples used
  for the construction of the groups of galaxies in the
  survey. $L_{eq}$ represents an estimate of the size of the
  equivalent cubic comoving box with the same volume as the sample of
  galaxies.}
\label{tab:tab1}
\end{table}


\section{Results}
\label{sec:results}

Tables \ref{tab:tab2} and \ref{tab:tab3} summarize the results of the
group finder applied to the four volume limited samples built from the
SDSS-DR7. There, richness, halo mass, stellar mass and luminosity
limits (minimum and maximum values per sample of galaxies) are shown.

In the next sections we study in detail the stellar mass content, halo
masses, luminosities and the richness of the groups identified in the
data with the implementation of the method presented in the previous
section.

\begin{table*}
\begin{center}
\begin{tabular}{|c|c|c|c|c|c|c|c|}\hline \hline

  Name   & $N_h$    & $\log_{10}(M_{min})$ & $\log_{10}(M_{lim})$ & $N_{gr}(N=1)$ &  $N_{gr}(N=2)$ & $N_{gr}(N=3)$ & $N_{gr}(N \ge 4)$ \\\hline
  Mr-18  & 38268    & 11.2                 & 14.7                 & 33863         &  2746          &   691         &  968         \\
  Mr-19  & 85222    & 11.5                 & 14.9                 & 76220         &  5568          &   1495        &  1939        \\
  Mr-20  & 128975   & 11.9                 & 15.1                 & 116620        &  7878          &   2077        &  2400        \\
  Mr-21  & 87040    & 12.7                 & 15.2                 & 80881         &  4344          &   996         &  819         \\

\hline
\hline
\end{tabular}
\end{center}
\caption{Table summarizing the properties of the groups for the four
  different samples. $N_h$ shows the final number of haloes in the
  sample. $M_{min}$ and $M_{max}$ are the minimum and maximum halo
  mass in the sample in units of \hMsun and $N_{gr}(N=1)$,
  $N_{gr}(N=2)$, $N_{gr}(N=3)$ and $N_{gr}(N \ge 4)$ show the number
  of groups with one, two, three and more than four members.}
\label{tab:tab2}
\end{table*}

\begin{table*}
\begin{center}
\begin{tabular}{|c|c|c|c|c|c|c|c|}\hline \hline

  Name   & $N_h$    & $\log_{10}(M_{stellar}^{min})$ & $\log_{10}(M_{stellar}^{lim})$ & $\log_{10}(L_{ch}^{min})$ &  $\log_{10}(L_{ch}^{max})$ & $R_{zs}^{min}$    & $R_{zs}^{max}$     \\\hline
  Mr-18  &  38268   & 8.9                            & 12.7                           & 9.1                       &  12.2                      &  0.9              &  11.6              \\
  Mr-19  &  85222   & 9.4                            & 12.6                           & 9.5                       &  12.1                      &  1.1              &  14.1              \\
  Mr-20  &  128975  & 9.8                            & 12.7                           & 9.9                       &  12.1                      &  1.6              &  15.2              \\
  Mr-21  &  87040   & 10.3                           & 12.5                           & 10.3                      &  11.8                      &  2.7              &  17.3              \\

\hline
\hline
\end{tabular}
\end{center}
\caption{Table summarizing the properties of the groups for the four
  different samples. $M_{stellar}^{min}$ and $M_{stellar}^{lim}$ show
  the maximum and minimum group stellar mass in each sample in units
  of \hMsun. $L_{ch}^{min}$ and $L_{ch}^{max}$ are the minimum and
  maximum group characteristic luminosity in units of \hLsun and
  $R_{zs}^{min}$ and $R_{zs}^{max}$ are the minimum and maximum
  ellipsoidal radius of search along the line of sight in \hMpc.}
\label{tab:tab3}
\end{table*}

\subsection{Stellar mass}
\label{sec:Mstellar}

As a first check, we verify that the stellar mass assignment in
galaxies produces reasonable results. Figure \ref{fig:StellasMassFunc}
shows the stellar mass function for galaxies and groups for our four
volume limited samples. Note that our data only goes until $\sim
10^{9.5}$\hMsun due to the high magnitude cut imposed to build the
volume limited samples. To make this comparison we have estimated the
mass function as the number density of galaxies with stellar mass $M$
in the range between $M$ and $M+dM$, where the stellar mass is the
result of the estimate from eqs. \ref{eq:MstellarGal} and
\ref{eq:Mstellar} weighted by the completeness of the survey at the
position of the galaxy, all divided by the total volume of the sample
of galaxies.

\begin{figure}
  \includegraphics[width=7.8cm,angle=270]{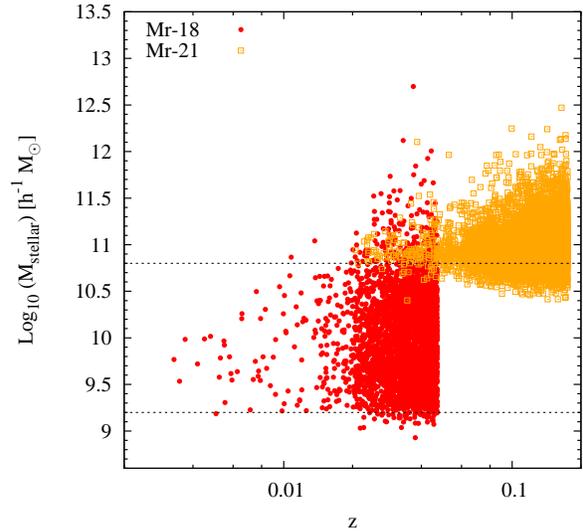}
  \caption{Group stellar masses as a function of redshift for the
    groups identified in two of the four samples. The horizontal lines
    indicate approximate masses where the set of haloes will be
    complete in stellar mass.}
  \label{fig:MStellar_z}
\end{figure}

For galaxies, our stellar mass function follows closely the mass
function presented in Baldry \etal (2008) in the range of masses
between $\sim 10^{9.5}$ and $\sim 10^{11.5}$\hMsun.  We can also see
in Figure \ref{fig:StellasMassFunc} that for masses below $\sim
10^{11.5}$ \hMsun the stellar mass function for groups have roughly
the same behaviour as the stellar mass function for individual
galaxies. For masses larger than $10^{11.5}$ the abundance of
stellar-massive objects is larger, but it is almost the same for the
three samples with the lower luminosity cuts. As expected, the number
density of groups with a given stellar mass is lower for the sample
associated with the high luminosity cut. This behaviour implies that
the contribution from galaxies with absolute magnitudes below
$M_r=-20$ to the stellar mass is important in characterizing the
baryonic content of massive groups. Also, as will be seen later in the
analysis of the richness, the groups with stellar masses below
$\sim10^{11.5}$\hMsun are associated to groups with less than $\sim10$
members. This low number of members, and therefore, low total stellar
mass, is responsible for the agreement between the stellar mass
function of individual galaxies and groups at low masses.

Since we assign masses to groups using the ranking of luminosities,
because for construction the samples are complete in luminosity, the
group stellar masses are not necessarily complete in each volume
limited sample. That is shown in Figure \ref{fig:MStellar_z}, where we
show the distribution of stellar masses for groups as a function of
redshift. The only sample that is almost complete in stellar mass is
Mr-18, the other three samples are incomplete due to the lack of
contributions to the stellar mass from the low mass galaxies not
included in each sample.

\begin{figure}
  \includegraphics[width=7.8cm,angle=270]{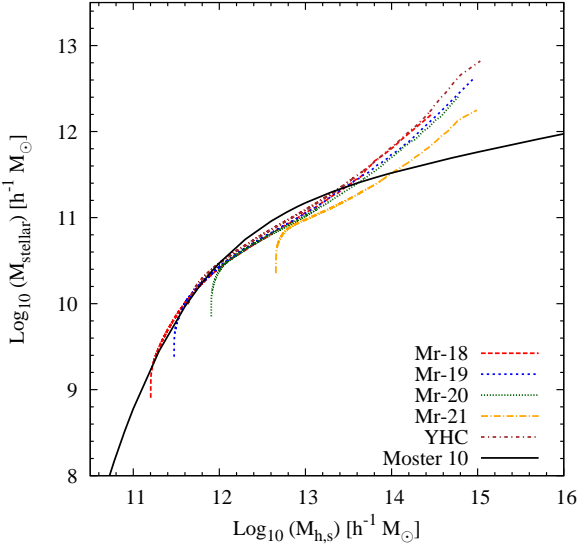}
  \caption{Stellar mass - halo mass relation for groups in the
    survey. Halo masses have been assigned using the ranking of group
    stellar mass. Filled-cyan points show the data from Yang \etal
    (2007) and the dashed line shows the prediction from Moster \etal
    (2010).}
  \label{fig:Mstellar_MhaloL18}
\end{figure}

Figure \ref{fig:Mstellar_MhaloL18} shows the stellar mass as a
function of halo mass for the four samples. For comparison we also
plot the results from the halo catalog of Yang \etal (2007) and the
fitting formula from Moster \etal (2010). Halo masses have been
computed through the ranking of the group stellar masses. Note however
that this mass assignment is technically incorrect if one uses samples
that are not complete in stellar mass. We have made comparisons
between the mass assignment using complete samples in stellar mass and
our final incomplete samples, and since the incompleteness affects a
small fraction of groups at the low mass end, the average results are
comparable. However, we will not assume this halo mass to be a final
reliable quantity. We keep it for completeness and to help us to
evaluate the performance of the halo mass assignment.

Here we see again the effect of the underestimation of the stellar
mass for the groups in the high luminosity sample, which in this case
leads to the underestimation of the stellar masses of groups hosted in
haloes with high masses. The small tails seen at the low halo mass end
are associated with the incompleteness of the samples in stellar mass
shown in Figure \ref{fig:MStellar_z}. We have tested it, and complete
samples in stellar mass do not present such a tails. From this figure
we can see, first, that there is agreement between the stellar and
halo mass among our four volume limited samples, as well as with the
halo catalog from Yang \etal (2007). Second, we see that our samples
also follow the expected stellar-halo mass relation from Moster \etal
(2010) up to $\sim10^{13.4}$\hMsun, where both, our catalogs, and the
one from Yang \etal (2007) show an upturn in the observed stellar
mass. This upturn seems to be associated to the stellar mass of
groups, not for individual star forming objects, as is assumed in
Moster \etal (2010).

\begin{figure*}
  \includegraphics[width=7.8cm,angle=270]{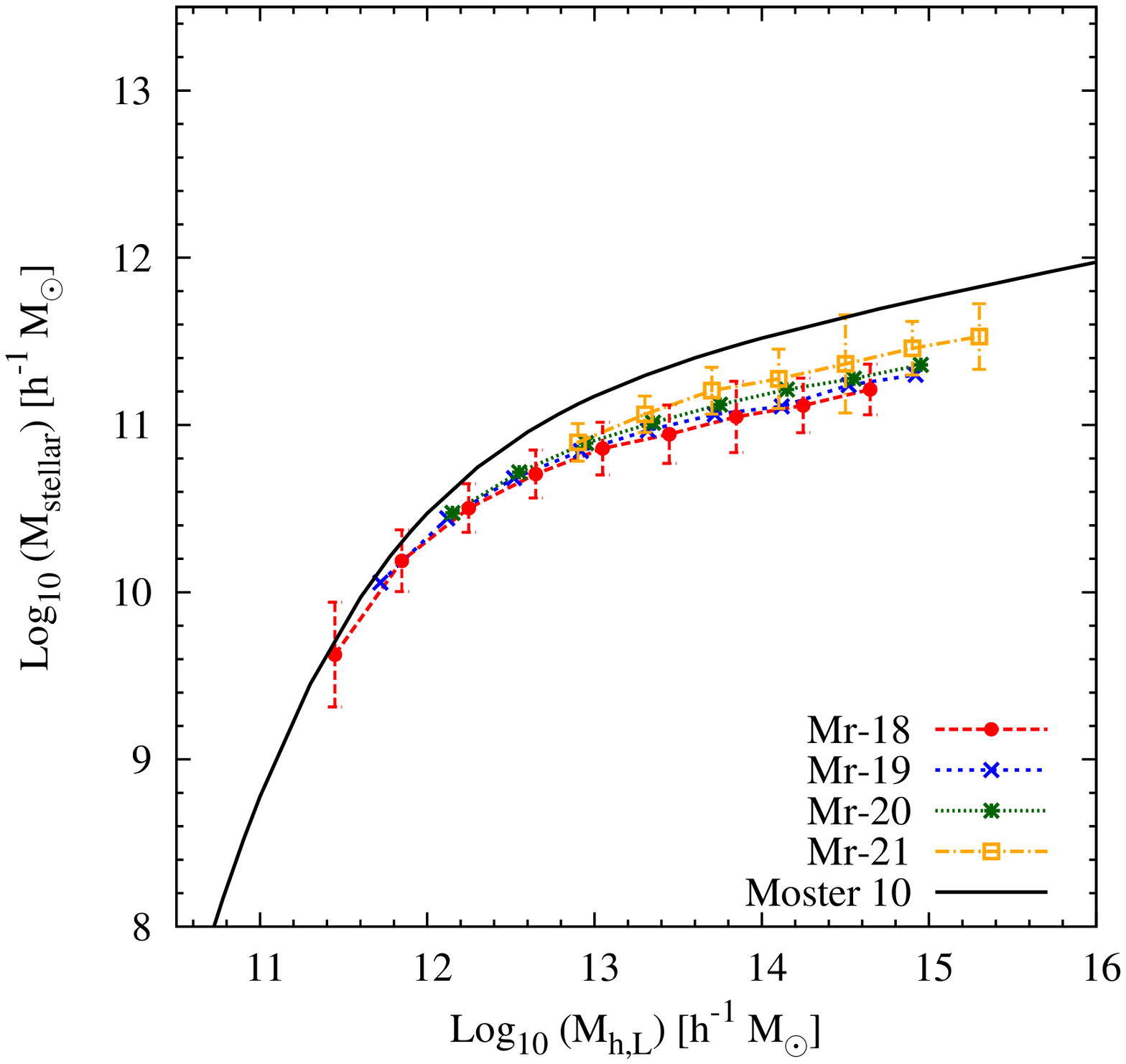}
  \includegraphics[width=7.8cm,angle=270]{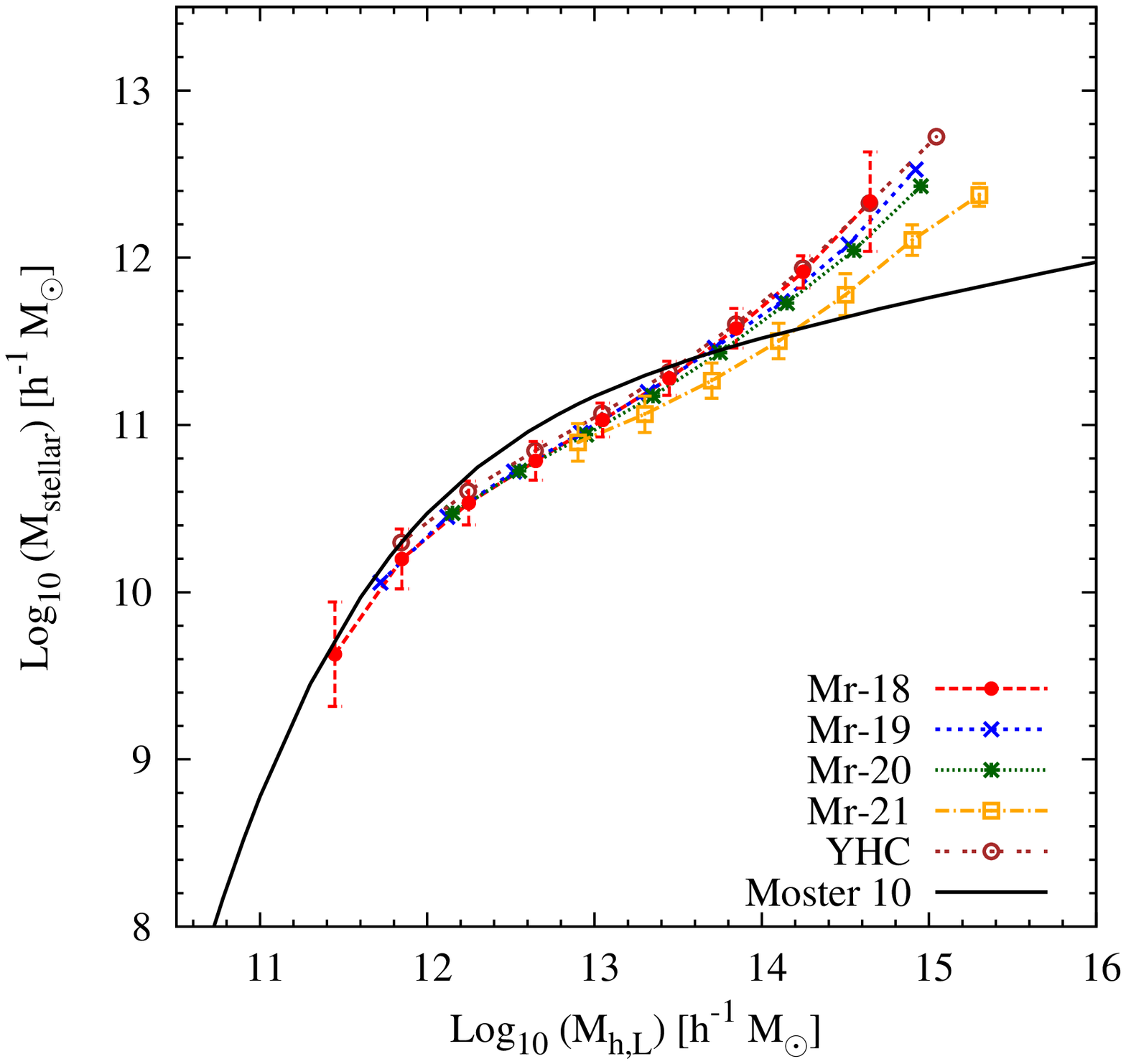}
  \caption{Median stellar mass of the central galaxy as a function of
    the halo mass. Halo masses are assigned using the ranking of the
    group characteristic luminosity. The frame at the left shows the
    relation using only the stellar mass in the central galaxy of the
    group while the frame at the right shows the the full stellar
    content in the group of galaxies computed from
    eq. \ref{eq:Mstellar}. Error bars indicate dispersion on the data
    in each bin.}
  \label{fig:MstellarMain_MhaloStellar}
\end{figure*}

Figure \ref{fig:MstellarMain_MhaloStellar} shows the stellar-halo mass
relation for groups and central galaxies. For these plots the halo
masses have been computed using the ranking on the group
characteristic luminosities. From those figures we see the origin of
the upturn in the stellar-halo mass relation for haloes more massive
than $\sim10^{13.4}$\hMsun. In figure
\ref{fig:MstellarMain_MhaloStellar} (left) we show the stellar-halo
mass relation, for the stellar mass of the central galaxy in the
group. Figure \ref{fig:MstellarMain_MhaloStellar} (right) shows the
same, but in this case the stellar mass includes the contribution from
all of the galaxies in the group. This confirms that the upturn in the
stellar-halo mass relation comes from the contribution of group
members in the stellar mass of the halo hosting it.

Note that for the sample Mr-21, the stellar-halo mass relation for
groups (Figure \ref{fig:MstellarMain_MhaloStellar}, right) is
different from the other three samples, while is the same for the
stellar-halo mass relation for the central galaxies (Figure
\ref{fig:MstellarMain_MhaloStellar}, left). This is the result of the
low group richness in this sample due to the absence of low luminosity
galaxies in Mr-21.

We can see also in Figure \ref{fig:MstellarMain_MhaloStellar} (left)
that the stellar-halo mass relation we obtain follows closely the
predictions from Moster \etal (2010), but differences appears and
increase for high halo masses. We see that at high halo masses our
data underestimates the stellar mass compared to the prediction. It is
difficult to find a reason for this discrepancy. If it were a problem
with the assignment of stellar mass to galaxies
(eq. \ref{eq:MstellarGal}), we should not be able to get the agreement
with the stellar mass function. If it where a problem with the halo
mass assignment, resulting from the estimation of $M_{min}$ and
$M_{lim}$, one would move our data points horizontally. That is,
changing $M_{min}$ and $M_{lim}$ in the same amount by modifying the
physical volume of the sample would solve the issue at the high mass
end, but will introduce a stronger disagreement at the low mass
end. Almost nothing will happen if we fix $M_{min}$ and make $M_{lim}$
smaller. Therefore we have no explanation for the differences.


\subsection{Characteristic luminosity}

\begin{figure}
  \includegraphics[width=7.8cm,angle=270]{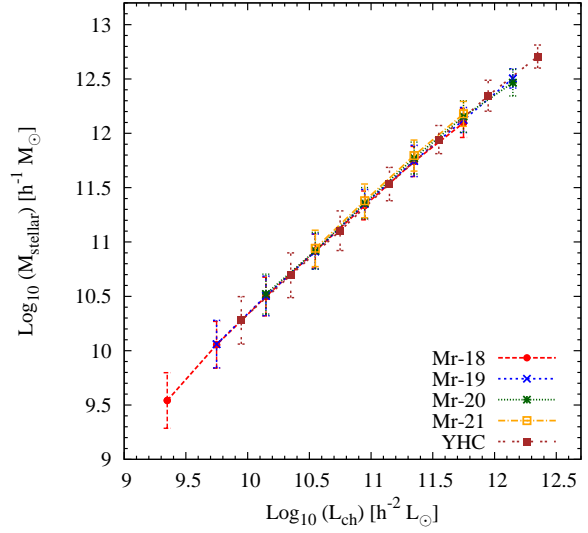}
  \caption{Median stellar mass as a function of the group
    luminosity. Error bars indicate dispersion on the data in each
    bin.}
  \label{fig:L18_Mstellar}
\end{figure}

Due to the volume limited nature of the samples, one of the most
important quantities for our catalog is the group luminosity. Figure
\ref{fig:L18_Mstellar} shows the group stellar mass as a function of
the group luminosity. Again, each colour shows the results for each of
the sample catalogs while the cyan points shows the comparison with
YHC. Note the scatter in stellar mass for a given halo
luminosity. This is partly due to the procedure used in the estimation
of the stellar masses of the galaxies in the galaxy sample, but must
also be due to a component of intrinsic scatter associated to
it. Clearly the scatter decreases as a function of group luminosity,
and we can see here again the presence of a few groups with very
massive stellar components, that seems to be in agreement also with
the results shown in YHC.

\begin{figure}
  \includegraphics[width=7.8cm,angle=270]{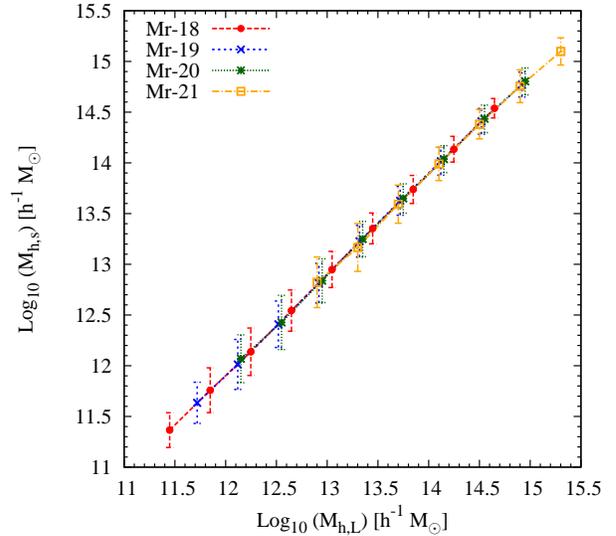}
  \caption{Median halo masses estimated using the ranking on the
    stellar masses as a function of the halo masses computed using the
    ranking of halo luminosities. Error bars indicate dispersion on
    the data in each bin.}
  \label{fig:MhaloLr18_MhaloStellar}
\end{figure}

\begin{figure*}
  \includegraphics[width=7.8cm,angle=270]{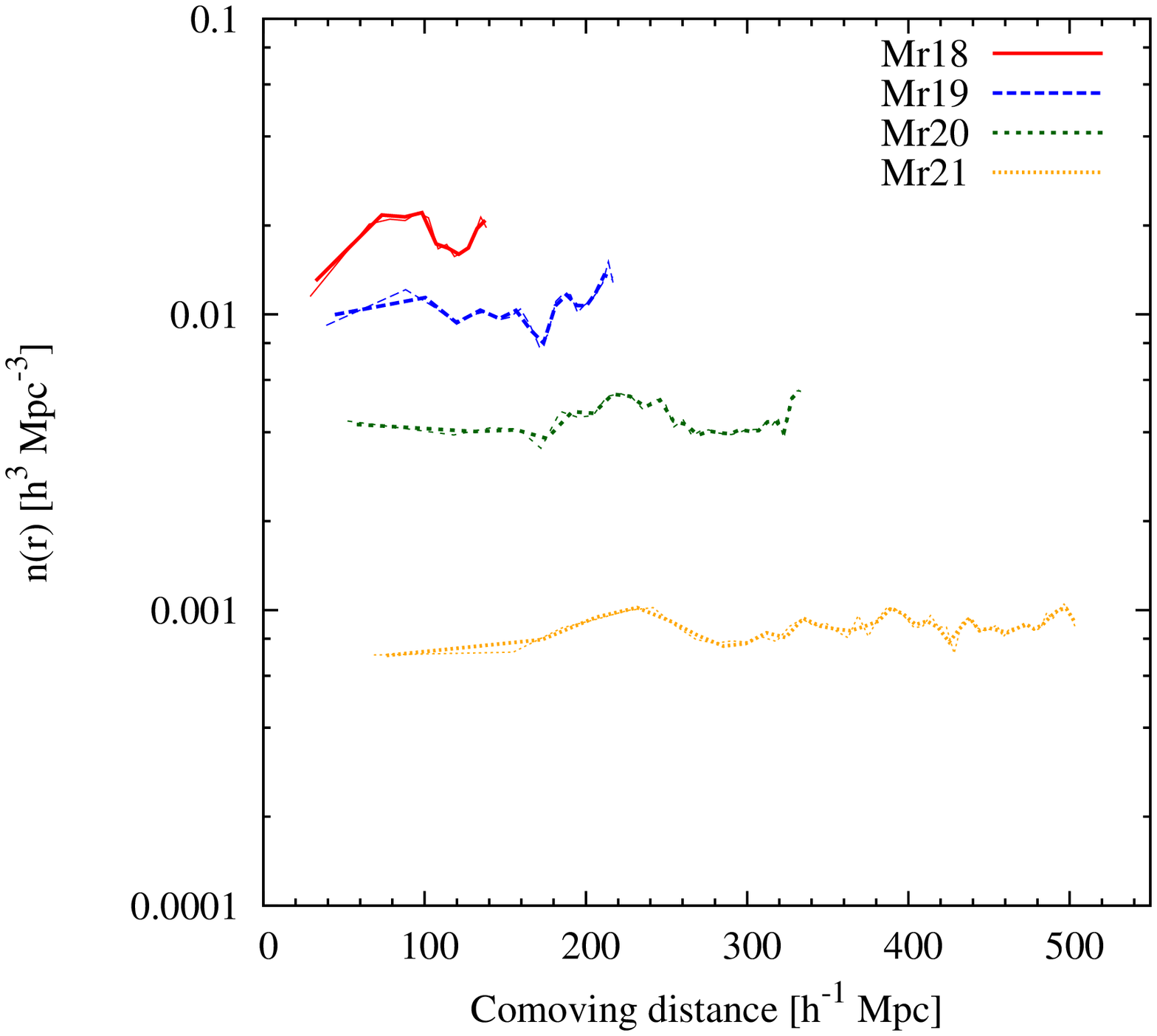}
  \includegraphics[width=7.8cm,angle=270]{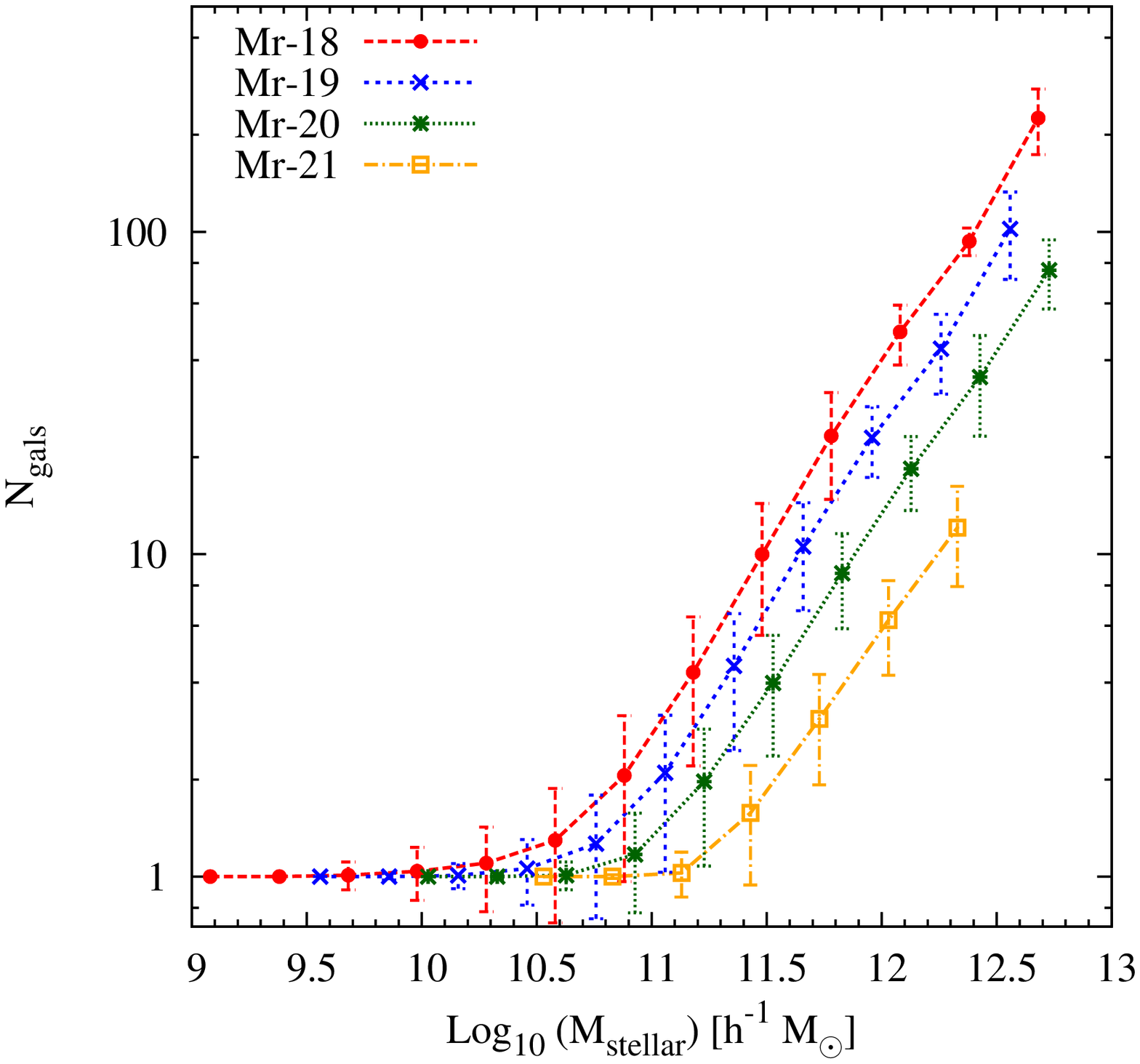}\\
  \includegraphics[width=7.8cm,angle=270]{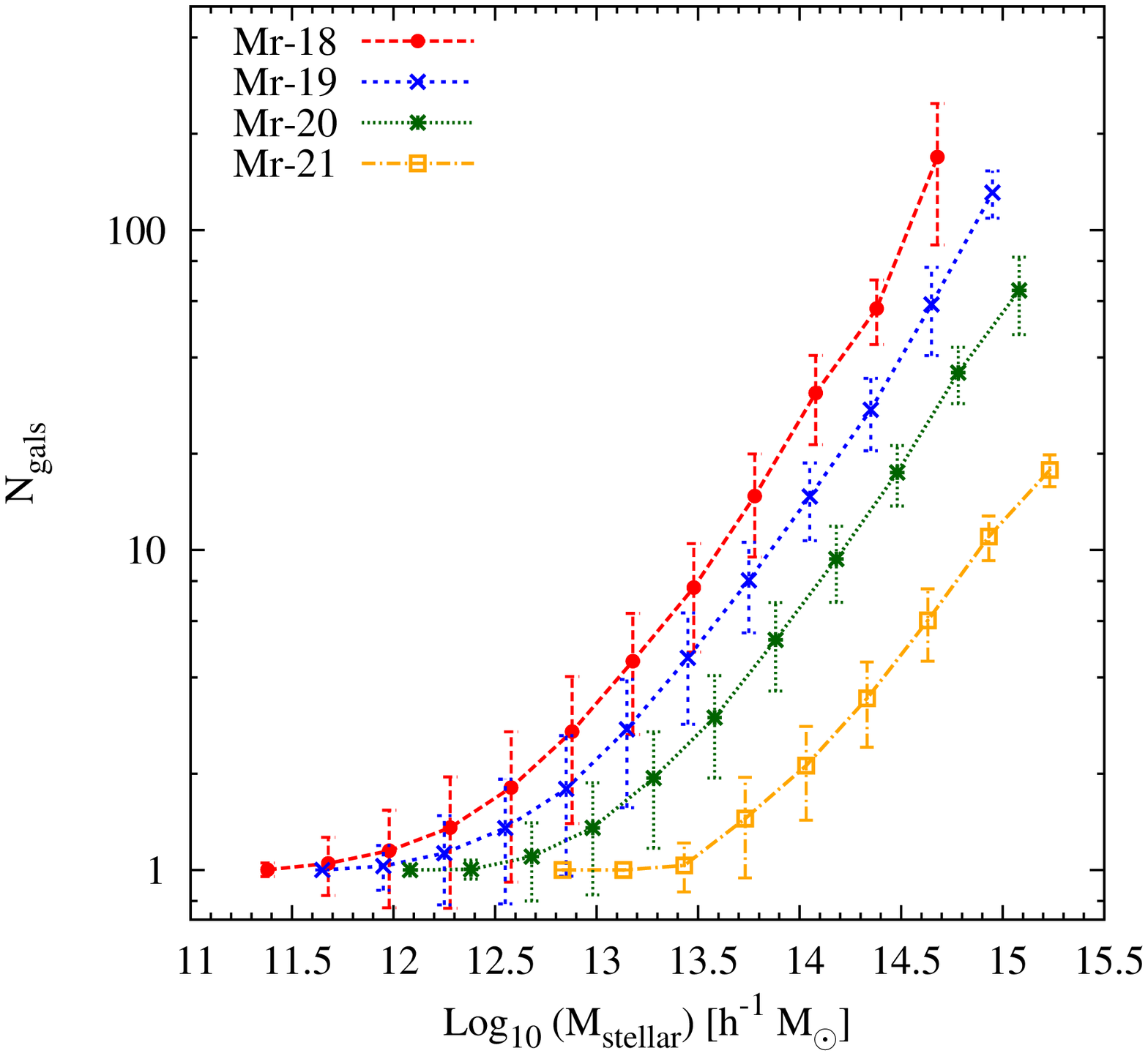}
  \includegraphics[width=7.8cm,angle=270]{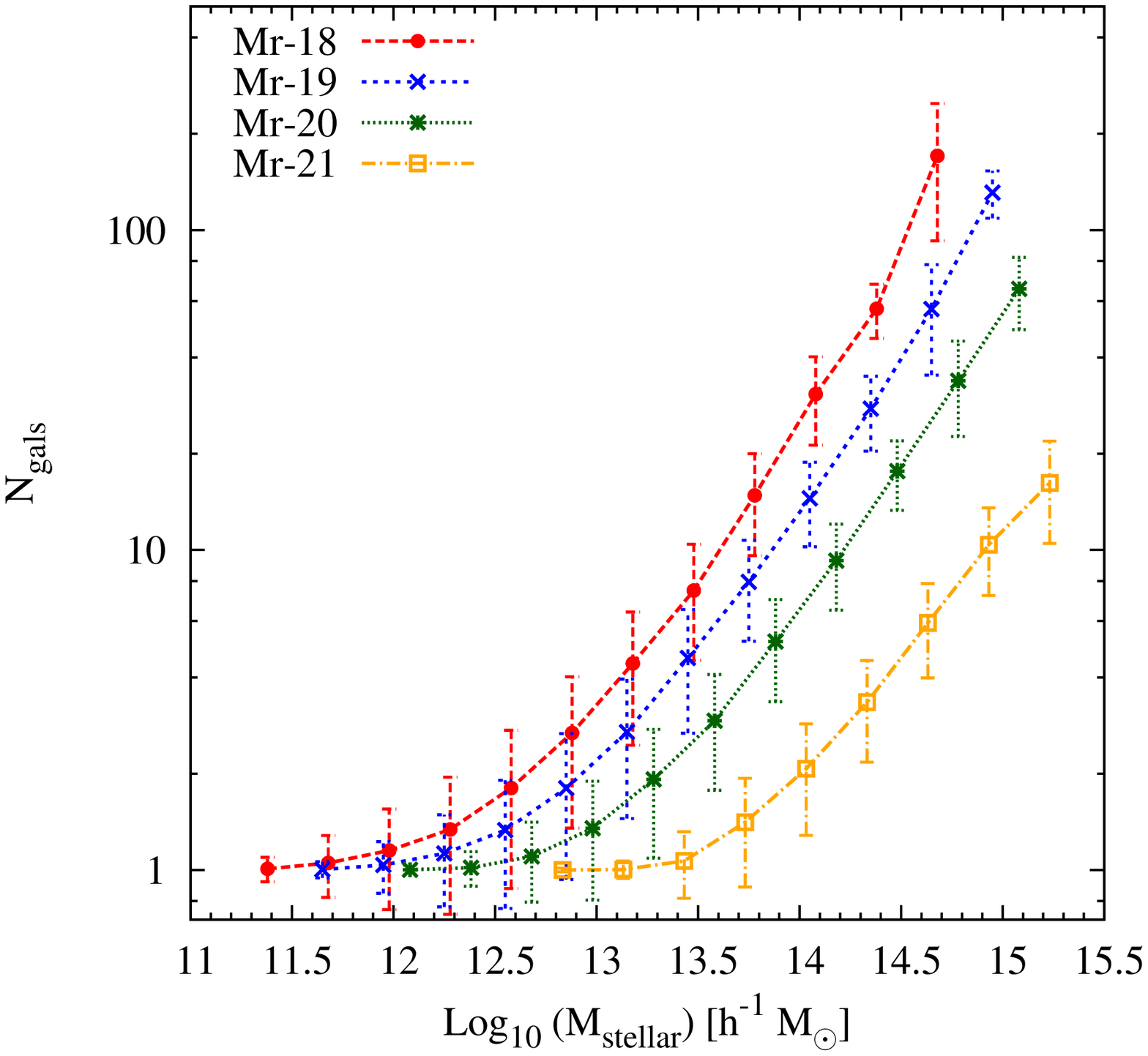}
  \caption{Mean number density of groups in the different samples as
    function of distance (upper left). Richness of groups as a
    function of group stellar mass (upper right), and halo mass
    estimated using the ranking of luminosities (lower left) and
    stellar masses (lower right). Solid lines show the mean value of
    the richness as a function of mass while the error bars indicate
    dispersion on the data in each bin.}
  \label{fig:Richness}
\end{figure*}

As a comparison, Figure \ref{fig:MhaloLr18_MhaloStellar} shows the
halo mass as computed from the ranking of the stellar masses
($M_{h,s}$) as a function of the halo mass computed from the ranking
of the group luminosities ($M_{h,L}$). Note first that there is a
scatter in the estimation of the masses for both methods, and the
scatter increases (for the same sample) at low halo masses. Although
there is a natural scatter in the distribution of stellar masses at a
given characteristic luminosity, this effect at low halo masses is
partly a consequence of the incompleteness in stellar mass of the
groups. This can be verified if one considers the small scatter for
the sample Mr-18, which is the sample close to be complete in stellar
mass. Interestingly, one can see that in the mean, both methods
provide the same mass assignment, the slope of the relation is the
same for all volume limited samples, which means that the estimated
halo masses are robust. Note that this effect is independent on the
incompleteness in stellar mass, and the already mentioned failure of
the sample Mr-21 to reproduce the stellar mass function. That both
methods produce, in the mean, the same halo mass is because the
implementation of the method in volume limited samples that influences
all groups by same amount.  This means that in general, if there is an
underestimation of stellar content in a group, all groups in the same
sample are missing the same fraction of mass, and at the end, the more
massive and luminous groups still are the most massive and luminous
ones independent on the magnitude cut. In that way, the ranking and
halo mass assignment is not affected by the absence of low mass-low
luminosity galaxies in the samples with a high luminosity cut.


\subsection{Richness}

Figure \ref{fig:Richness} shows the mean comoving density of groups as
a function of redshift (top left). The richness of groups as a
function of group stellar mass (top right) and halo mass estimated
using the two different approaches, the ranking of the halo
luminosities (bottom left) and the ranking of the stellar masses
(bottom right). The lines with symbols in those plots show the mean
richness as a function of mass computed in mass bins of 0.3dex.

The mean density is computed as described in Mu\~noz-Cuartas \etal
(2011). Using a Delaunay tessellation we compute the mean number
density of groups as a function of distance from the observer. As it
can be seen in Figure \ref{fig:Richness}, the mean number density of
groups is almost constant for all samples. Again, the different
normalization is due to the different abundance of groups identified
in the different galaxy samples due to the different luminosity
cuts. The thick lines in figure \ref{fig:Richness} shows the mean
number density estimated using radial bins of 14\hMpc width while the
thin lines show the same computed in radial bins of 10\hMpc.

We can see in Figure \ref{fig:Richness} that in all samples, the
groups with the largest stellar mass and the largest halo mass are the
richer ones. We can see in particular, that the group with the largest
stellar mass is resolved by many galaxies in the low luminosity
samples, but is resolved with only one galaxy in the volume limited
sample with the highest luminosity cut. Besides that, we see that
there is a direct proportion between stellar mass and richness of
groups. The same behaviour is observed in the lower panels that show
the richness as a function of halo mass.


\subsection{Effects of varying the mass function}

While it is true that our method does not require any parameter, there
are a couple of assumptions that can affect the results of the group
finding.

The first is the assumption of the validity of
eq. \ref{eq:MstellarGal} to compute the stellar masses of individual
galaxies. We have already tested the performance of such an
approximation (see section \ref{sec:Mstellar}) through comparisons
with the stellar mass function. From the agreement we observe between
the expected and our estimated stellar mass function we assume there
is no major influence on our results from this approximation.

The second assumption concerns the mass assignment. First, we assume
that the halo virial mass is equivalent to the mass we draw from the
halo mass function. It has been shown (White 2000) that there are
deviations between our definition of virial mass and the halo mass
estimated through different criteria (FoF mass, M200, etc.), but these
deviations are not larger than a factor of two. This makes our mass
estimates to be within the scatter of the mass distribution.

Finally, the shape of the assumed mass function can have an impact on
the masses of the haloes associated to each group of galaxies. Since
it is known that different fitting functions (Sheth \& Tormen 2002,
Warren \ etal 2006, Reed \etal 2007, Tinker \etal 2008) can give
slightly different mass function, the procedure of reconstruction may
be affected by this factor.

\begin{figure}
  \includegraphics[width=7.8cm,angle=270]{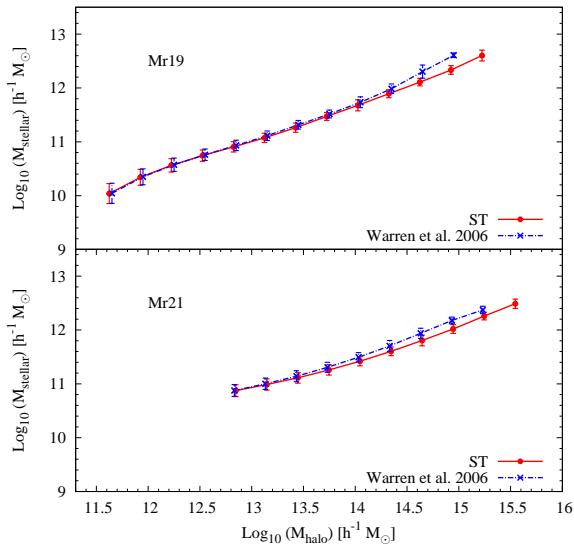}
  \caption{Median stellar mass as a function of halo mass for two of
    the samples, using two different mass functions, Sheth \& Tormen
    (2002) and Warren \etal (2006). Halo masses are computed through
    the ranking of halo luminosities. Error bars indicate dispersion
    on the data in each bin.}
  \label{fig:Mstellar_MhaloL18_ST_W.ps}
\end{figure}

Figure \ref{fig:Mstellar_MhaloL18_ST_W.ps} shows the final group
stellar masses as a function of halo mass for groups of galaxies in
two different volume limited samples, Mr-19 and Mr-21. In both cases
the halo masses have been computed through the ranking of halo
luminosities using the mass function from Sheth \& Tormen (2002) and
Warren \etal (2006). We can see that the halo masses assigned using
the mass function of Warren \etal (2006) are systematically lower as
compared with the samples of haloes with masses assigned from the
Sheth-Tormen mass function. This differences should also affect the
stellar masses, abundance of haloes and richness of groups, since the
halo mass controls the virial radius and circular velocity, which are
used to compute the adaptive linking lengths of the search. This
effect is stronger in the sample with the higher luminosity cut, and
for high masses (larger that $\sim10^{13.5}$\hMsun) due to the fact
that the differences between both mass functions are larger at the
high mass end.

Because of the adaptive modifications in the linking lengths, the two
halo catalogs are not identical, therefore we can not make a full
one-to-one comparison between the masses assigned using the two mass
functions.

However, we have seen that depending on the used mass function, one
can have differences in the total number of groups by an order of 2\%,
slight changes in the stellar masses for haloes more massive that
$\sim10^{11.5}$\hMsun by in average $\sim 0.1$ dex, and changes in the
maximum halo mass by around 0.4dex. Despite these differences, which
are observed mostly at the high mass end, the average statistical
differences obtained using the different mass functions are small when
one compares all the galaxy samples.

\subsection{Tests against mock catalogs}

One of the best ways to test the performance of the group finder is
using the results of semi-analytic methods of galaxy formation. For
a galaxy catalog built from a semi-analytic method we can compare the
results of our method against the original expected distribution of
groups and galaxies and their properties.

To do this we have used the galaxy catalogs (De Lucia \& Blaizot 2007,
Croton \etal 2006) built from the millennium simulation (Springel
\etal 2005). From the catalogs we have extracted a cubic subvolume of
200\hMpc side length. From this volume we built two samples using two
different magnitude cuts in $M_r$ of -18 and -19 to resemble our
samples Mr-18 and Mr-19. Since the results are comparable for both
samples, in the following we will present results only for the
subsample with magnitude cut of -19.

For this sample we used the peculiar velocities of the galaxies to
introduce redshift space distortions along the z-axis using the far
observer approximation. We have run our group finder in two versions
of the galaxy sample, one in real space and another in redshift
space. Running the group finder for the set of galaxies in real space
works as a control setup and allow us to identify the effects of the
different approximations of the method in our results. Running the
group finder in redshift space gives us information about the effect of
redshift space distortions and contamination.

\begin{figure}
  \includegraphics[width=7.8cm,angle=270]{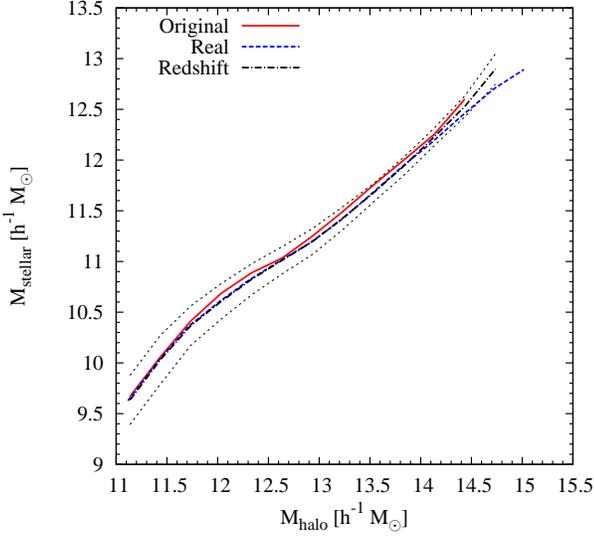}
  \caption{Median stellar mass - Halo mass relation for groups of
    galaxies in the mock catalog. The red solid line shows the values
    for original groups as obtained from the galaxy catalog. The blue
    dashed line show the groups identified using the group finder and
    the black dot-dashed line show the median value for the sample of
    galaxies in redshift space. The thin dashed line shows the the
    scatter on the data for the sample of galaxies in redshift space.}
  \label{fig:Mhalo_Mstellar_RealRedshift.ps}
\end{figure}

In Figure \ref{fig:Mhalo_Mstellar_RealRedshift.ps} we show the median
halo mass - stellar mass relation for groups in the mock galaxy
samples. Each line shows the relation for the original data from the
mock catalog (Original), the groups identified in the sample using our
group finder in real space (Real) and the groups identified with our
method on the sample in redshift space (Redshift). In the figure, the
thin-dashed line shows the scatter on the data for the sample in
redshift space. One of the advantages of using mock catalogs is that
it provides not only the information of the galaxy properties, but
also provides the link to the properties of the host haloes. This
allows us to check the reliability of the halo mass assignment. As it
is shown in Figure \ref{fig:Mhalo_Mstellar_RealRedshift.ps}, the
halo mass assignment produces halo masses that show a very good
agreement between the two samples and the original data. We can see
that there are small deviations in the stellar-halo mass relation for
the groups identified in both, real and redshift space. This is partly
due to incompleteness in the galaxy and halo samples at low masses
(similar to those discussed in Figure \ref{fig:Mstellar_MhaloL18}). As
discussed in the previous section, another factor inducing small
differences is the halo mass function used to perform the group
finding, that might reproduce closely but not exactly the mass
function of the simulation. We find that our method assigns a larger
maximum mass to haloes in the high mass end. At the low mass end we
have some low mass haloes in the true halo catalog which can not be
regarded as individual haloes by our group finder and therefore are
merged with other haloes.

\begin{figure}
  \includegraphics[width=7.8cm,angle=270]{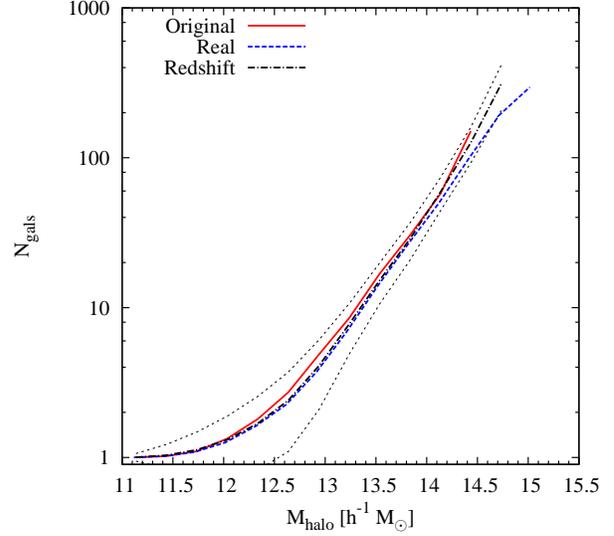}
  \caption{Mean richness of groups as a function of halo mass for the
    three samples, original, real space and redshift space. As in the
    previous figure, the thin lines show the scatter for the data of
    the sample of galaxies in redshift space.}
  \label{fig:Mhalo_Ngals_RealRedshift.ps}
\end{figure}

Despite the small differences at the low and high mass ends, the
median values of the stellar-halo mass relation show very good
agreement, even for the sample in redshift space. The difference is
well below the scatter of the data. A similar behaviour is observed in
Figure \ref{fig:Mhalo_Ngals_RealRedshift.ps} where we show the mean
richness of haloes as a function of halo mass. Again the differences
in the richness of haloes are smaller than the scatter in redshift
space, where the method produces the largest scatter.

\begin{figure}
  \includegraphics[width=7.8cm,angle=270]{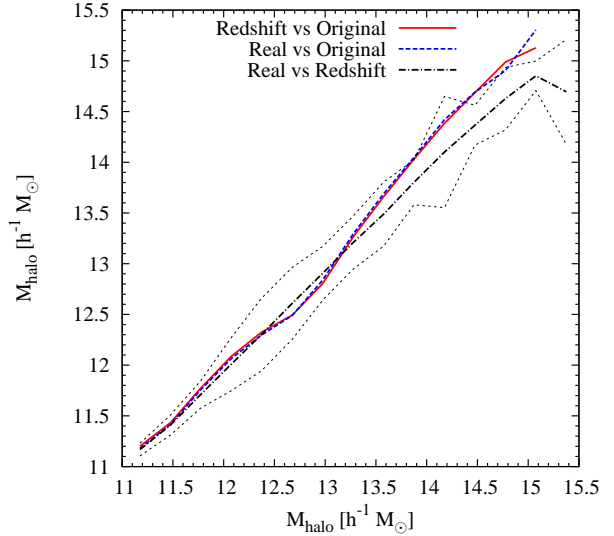}
  \caption{One to one comparison of the halo mass of haloes in the
    three samples. The solid red line shows the halo mass for the same
    haloes in the sample in redshift space against the halo mass in
    the original catalog. The blue dashed line shows the same for the
    groups in the sample in real space and the original catalog, and
    the black dot-dashed line shows the relation of the data in real
    and redshift space.}
  \label{fig:One2One_Mhalo.ps}
\end{figure}

In order to provide a closer comparison, we have made a cross check of
the halo masses for the groups of galaxies with the same central
galaxy among the different samples. In Figure
\ref{fig:One2One_Mhalo.ps} we show the median one-to-one halo mass
cross-check for the groups in the three samples. As it can be seen in
the Figure, the one-to-one comparison shows very good agreement for
the three samples, with the halo mass estimated from the group finder,
in real as well as in redshift space, being slightly larger for haloes
with masses above $\sim10^{13.5}$ \hMsun. From the figure it is
evident that it is not an effect of the performance of the group
finder in redshift space. The small difference of less than $\sim0.2$
dex between the original halo mass and the mass assigned by the group
finder comes mostly from the assumed mass function. As we have shown
in Figure \ref{fig:Mstellar_MhaloL18_ST_W.ps} it can influence the
estimated halo mass especially at the high mass end. From these tests,
we can conclude a good performance for the group finder. In particular
we can see that the halo mass assignment produces very reliable
results, with differences of the order of at most 0.2 dex. As
expected, the richness of the groups is the most sensible quantity,
and the effect is more noticeable at low halo mass, or equivalently
for low richness groups, where we see that at a given halo mass the
scatter increases, particularly for groups with richness below
three. The good agreement we obtain between the two samples in real
and redshift space with the original data from the simulation allows
us to conclude that the approximations used in the method to estimate
group-halo properties are reasonable, and that the scheme of group
finding we propose produces reliable results. In general we see these
results as a key point in favour of the quality and reliability of the
method and its results.

A more direct check on the quality of the performance of the group
finder and the quality of the group catalogs it produces is obtained
through the analysis of the purity, completeness and contamination by
interlopers in the groups identified from the mock catalog as
introduced in Yang \etal (2007). Here we summarise the procedure and
definitions.

Assume we have two group catalogs, one having the true groups, and the
other one having the groups identified by the group finder in redshift
space. We assume that two groups in the two catalogs are the same if
the ID of the central galaxy is the same. The group in the true
catalog has $N_t$ galaxies, while the group identified by the group
finder has $N_f$ galaxies. If $N_c$ represents the number of common
galaxies between the two groups, then we define the purity $f_p$, the
completeness $f_c$ and the contamination $f_i$ as

\begin{eqnarray}
f_p &\equiv& N_t/N_f, \nonumber\\ 
f_c &\equiv& N_c/N_t, \\
f_i &\equiv& (N_f-N_c)/N_t, \nonumber
\end{eqnarray}

\noindent
Note that with this definition, a perfect group finder will produce
groups with $f_p=1$, $f_c=1$ and $f_i=0$.

\begin{figure}
  \includegraphics[width=7.2cm,angle=270]{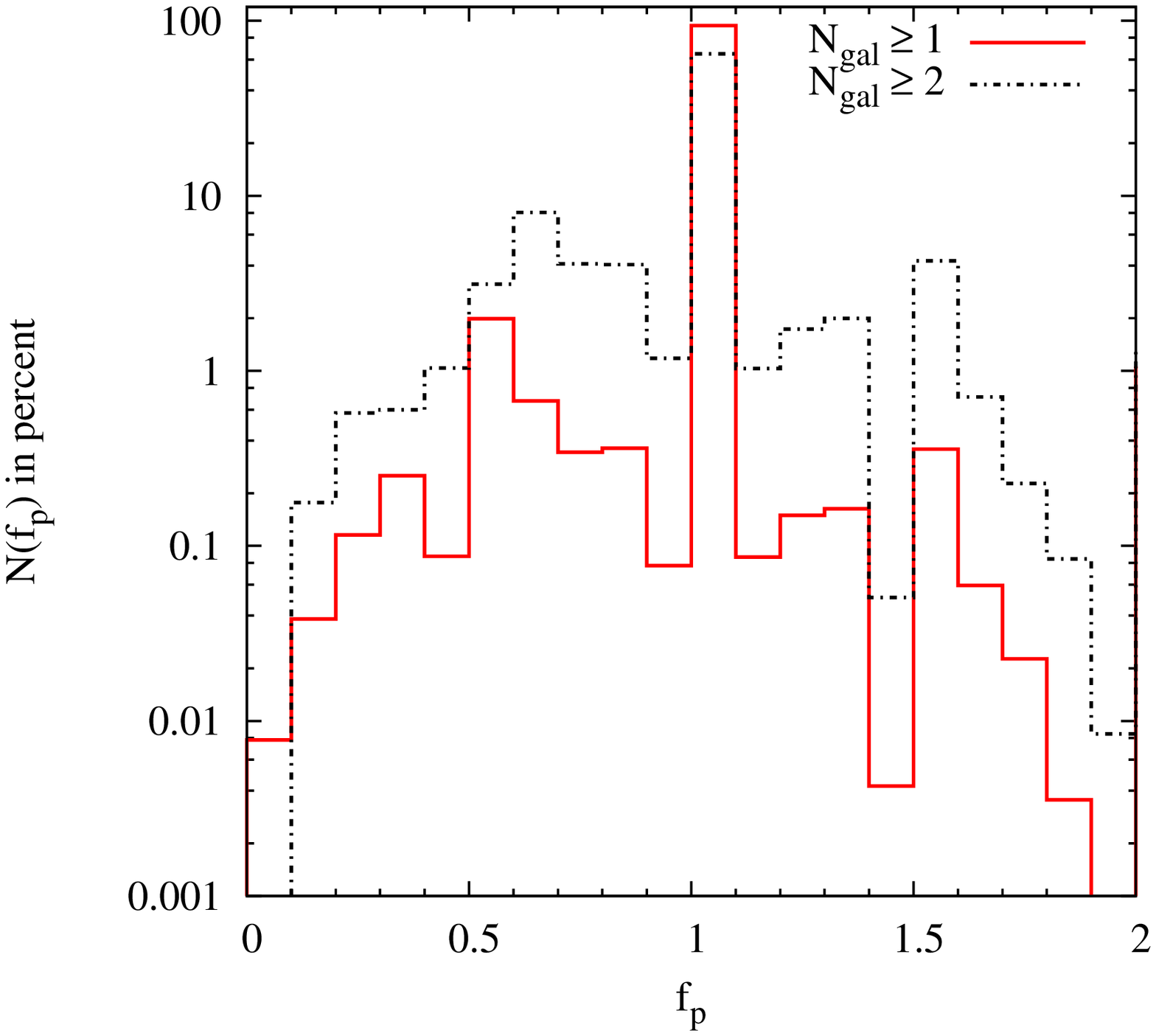}\\
  \includegraphics[width=7.2cm,angle=270]{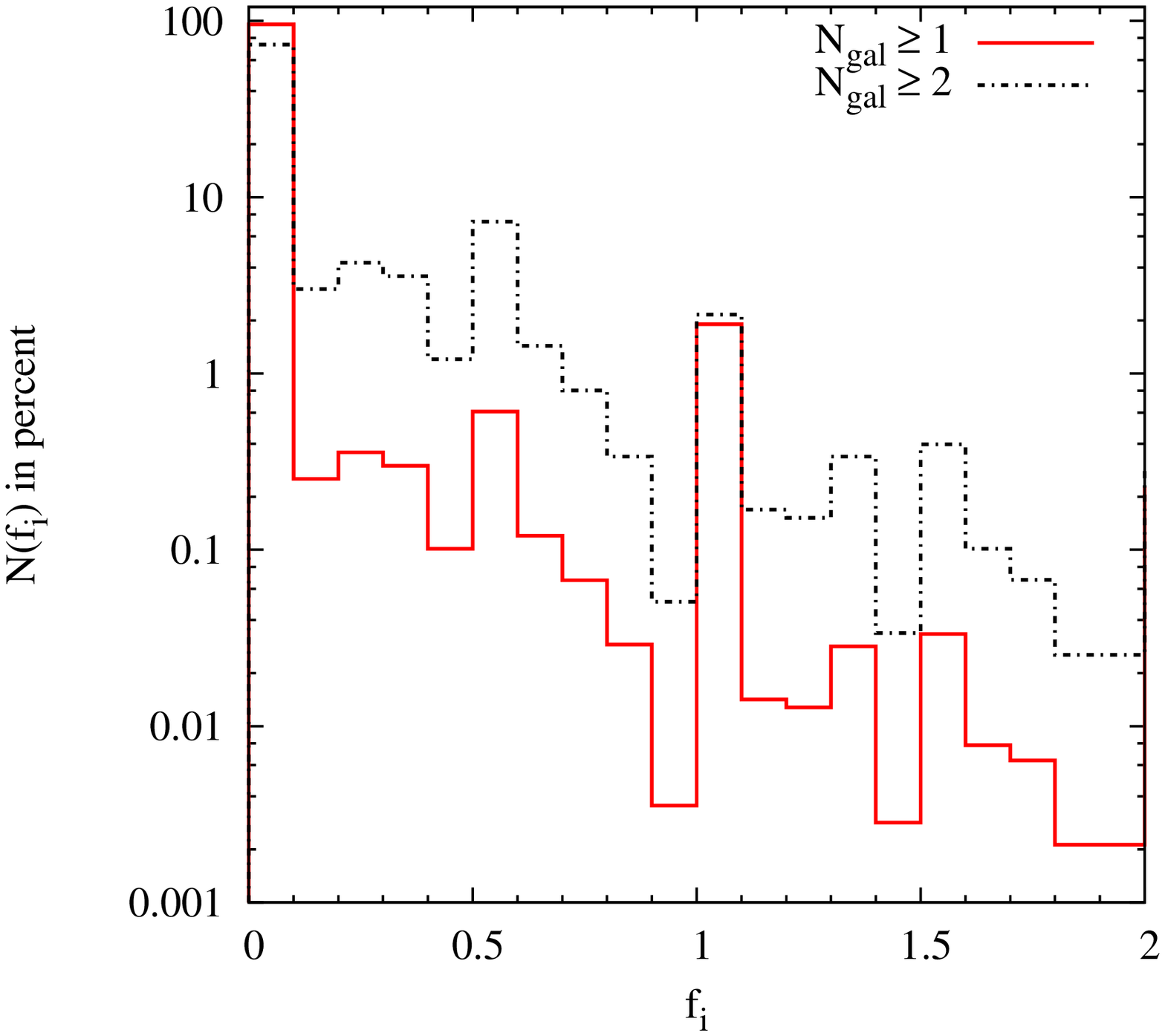}\\
  \includegraphics[width=7.2cm,angle=270]{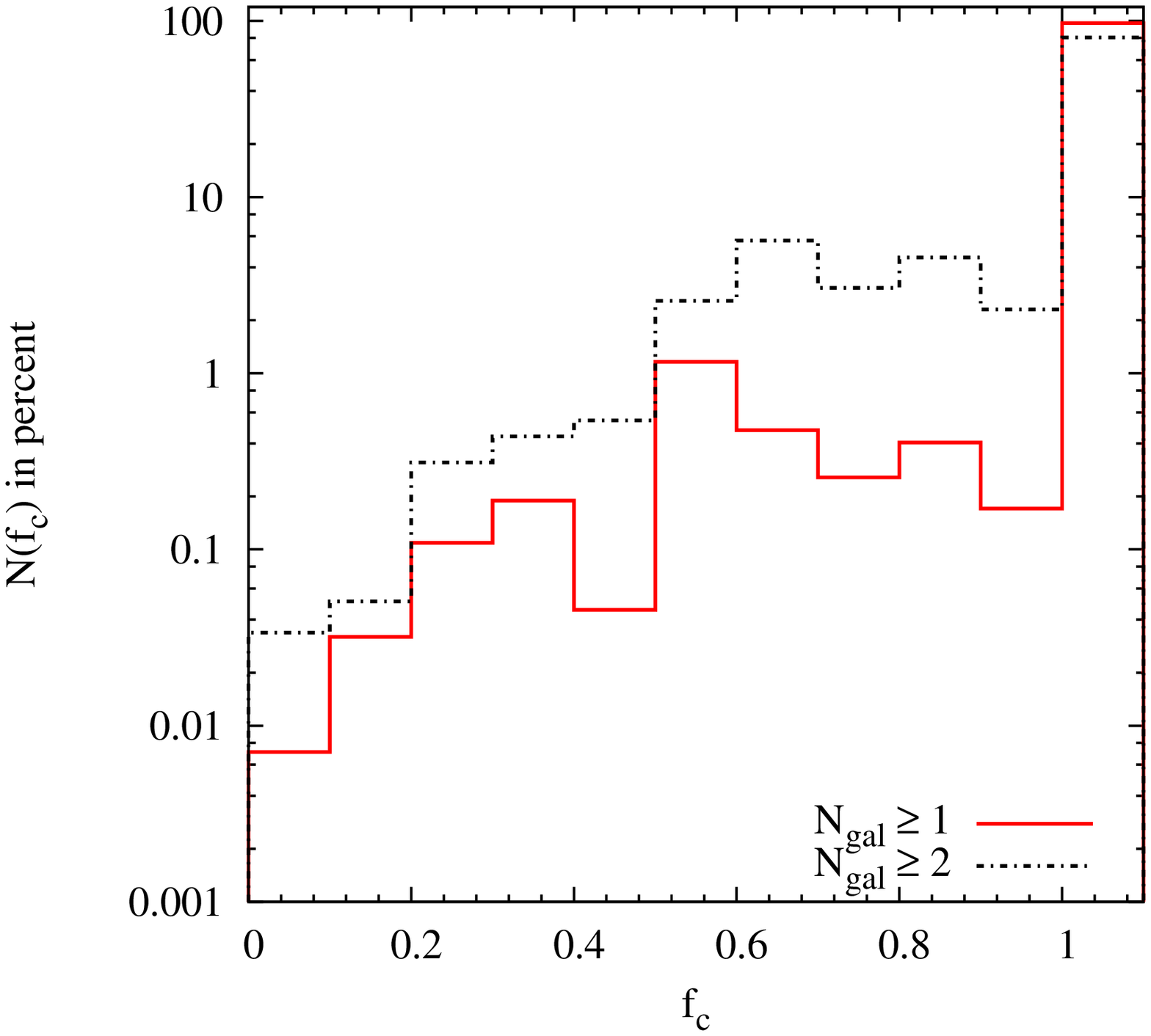}\\
  \caption{Distributions of purity (top), contamination by interlopers
    (middle) and completeness (bottom). The solid line shows the
    values for groups with at least two galaxies. As a comparison, the
    dashed line shows the values for groups with at least one galaxy.}
  \label{fig:Purity}
\end{figure}

In Figure \ref{fig:Purity} we show the distribution of values for the
purity, contamination and completeness in percents. The solid line
show the results for groups with at least one member, while the dashed
line show the results for groups with at least two members. As
mentioned in Yang \etal (2007), groups with only one member will have
$f_i=0$. To make a more precise analysis, it is necessary to account
for the effect of the groups with only one member separately. Not
including them hides the ability of the finder to identify real groups
with only one member. Including them without considering (at least as
a comparison) their effect on $f_c$, $f_i$ and $f_p$ could lead to a
overestimation of the quality of the group finder and the catalogs it
produces, as it is shown in Figure \ref{fig:Purity}.

In general, for groups with $N_{gals}\leq2$, we see that around 64\%
of the groups have a purity of $f_p=1$, 79\% have purity between 0.7
and 1.3 and that around 94\% of the groups have a purity between 0.5
and 1.5. On the other hand, $\sim80\%$ of the groups are hundred
percent complete ($f_c = 1$). 96\% of all groups have completeness
above 0.6 while $\sim87\%$ have completeness above 0.8. Finally, 73\%
of the groups have no contamination ($f_i=0$), while 92\% of them have
contamination lower than 0.5. In general these numbers indicate that
the performance of the method is very good, and it is in average
slightly better than the original method proposed in Yang \etal
(2007). This good performance, and the parameter-free nature of the
method are the most important features of the group finder.


\section{Summary and discussion}

We have presented a method for the identification of groups of
galaxies and associated dark matter haloes in galaxy redshift
surveys. We have applied the method to the data of the Seventh Data
Release of the SDSS. The method, that works like a FoF method, uses a
local and adaptive linking length that depends on the properties of
haloes. The most important property of the method is that it does not
depend on any parameter.

The method presented in this work is based on that introduced in Yang
\etal (2007), nevertheless differs from it in several points. In our
implementation of the method we do not need to make a first FoF
procedure in redshift space to start the iterations of the
groups. This avoids possible contamination by the choice of the
initial linking lengths. Also, differently from the assumption made in
Yang \etal (2007), we do not need to assume an initial value for the
mass-to-light relation of groups. Finally, our method uses a two
dimensional spheroid for the search of group members. This is
different from the implementation in Yang \etal (2007), where they use
a fixed FoF linking length for the transverse search and a
probabilistic approach for the redshift distribution of galaxies, that
at the end requires the use of another free parameter to fix the
density contrast defining the membership of galaxies in groups.

We have shown that the stellar mass assignment for individual galaxies
produces results that are in good agreement with the stellar mass
function of Baldry \etal (2008). We also have seen that the stellar
mass function of groups is more or less similar to the stellar mass
function of individual galaxies for stellar masses smaller than
$\sim10^{11.5}$\hMsun. For stellar masses above $10^{11.5}$\hMsun the
abundance of massive objects is larger for the groups than for the
individual galaxies. This behaviour is expected, since groups of
galaxies should have larger stellar masses than individual galaxies,
increasing the abundance of stellar massive objects.

An important conclusion from the analysis of the stellar mass function
of groups, that although trivial, is important to be quantified in
detail. It concerns the effect of the richness of groups on their
stellar content. We have seen in the stellar mass function for groups
and also in the stellar-halo mass relation, that the stellar content
of the volume limited sample built with the largest luminosity cut
(Mr-21) has lower values for the group stellar masses. We have seen
also that the richness of groups in this sample is the lowest one, due
to the high luminosity cut that reduces the number density of galaxies
in the sample. These results imply that most of the stellar content in
groups of galaxies comes from objects with absolute magnitudes $M_r$
larger than -19. This means that it is important to resolve the low
luminosity component of groups of galaxies to acquire detailed
information about their properties and specifically, its stellar
content.

We have also shown that the groups built in this work follow the
stellar-halo mass relation shown in Moster \etal (2010). We see that
at halo masses larger than $\sim10^{13.5}$\hMsun there is an increase
in the stellar mass of our groups, making groups hosted in these
massive haloes to have larger stellar mass content (larger than
$\sim10^{11.5}$\hMsun) than expected from the prediction of Moster
\etal (2010). We have shown that this deviation vanishes when the
stellar-halo mass is plotted using only the stellar mass of the
central galaxy in each group. Therefore the large abundance of high
stellar mass groups is due to the contribution of satellite galaxies
in rich groups.

We have also shown that the halo mass computed from the ranking of the
group luminosities $M_{h,L_{ch}}$ or from the ranking of the stellar
mass $M_{h,M_{s}}$ are in good agreement. Furthermore, as a
confirmation of the robustness of the mass assignment, we see that the
halo mass assignment produces results that are in agreement among all
our four samples.

Tests against galaxy catalogs from semi-analytic methods have proven
the good performance of the method, showing that we can recover with
high precision the properties of the groups of galaxies and haloes in
the catalog. Specifically, a one-to-one comparison has shown that the
halo masses are in good agreement, but are slightly off by at most 0.2
dex for halo masses larger than $10^{13.5}$\hMsun. On the other hand,
purity, completeness and contamination indicators show a good
performance of the method, and in general, show to be better than the
ones presented by Yang \etal (2007). The validation of the method with
galaxy catalogs provides us with strong evidence in favour of the
convenience of the use of this method for the identification of galaxy
groups residing in the same dark matter halo.

We have found that the only possible factor that can influence the
final results of our group-halo identification is the selection of the
mass function. To test the influence of this factor on our results, we
have compared the results of the identification of groups in two of
the samples. For them we have computed the halo masses from two
different mass functions. We have seen that depending on the used mass
function one can have differences in the total number of groups of
about 2\%. We observe a slightly changes in the stellar masses of
groups (in average on $\sim 0.1$ dex) hosted in haloes more massive
that $\sim10^{14.5}$\hMsun. We see also changes in the maximum halo
masses of around $\sim0.4$ dex. Despite these differences, that are
observed only at the very high halo masses, overall differences
obtained using different mass functions are minor.

Finally we have shown the richness of our groups as a function of
different group-halo properties. First we have seen that we can
recover a constant mean number density of groups in space, as it
should be the case for volume limited samples. We have seen that
indeed the volume limited nature of the samples affects the richness
of the groups at a given halo property, groups in samples with a high
luminosity cut are naturally less rich. This has an important
effect in the estimation of the stellar mass of groups but does not
have implications on the halo mass assignment, and therefore in the
procedure of group identification.

\section*{Acknowledgments}

J.C.M. wants to give thanks to German Science Foundation (grant MU
1020/6-4). The authors wants to thanks Sebastian Nuza for his useful
comments on the manuscript. The Millennium Simulation databases used
in this paper and the web application providing online access to them
were constructed as part of the activities of the German Astrophysical
Virtual Observatory. Funding for the Sloan Digital Sky Survey (SDSS)
has been provided by the Alfred P. Sloan Foundation, the Participating
Institutions, the National Aeronautics and Space Administration, the
National Science Foundation, the U.S. Department of Energy, the
Japanese Monbukagakusho, and the Max Planck Society. The SDSS Web site
is http://www.sdss.org/. The SDSS is managed by the Astrophysical
Research Consortium (ARC) for the Participating Institutions. The
Participating Institutions are The University of Chicago, Fermilab,
the Institute for Advanced Study, the Japan Participation Group, The
Johns Hopkins University, Los Alamos National Laboratory, the
Max-Planck-Institute for Astronomy (MPIA), the Max-Planck-Institute
for Astrophysics (MPA), New Mexico State University, University of
Pittsburgh, Princeton University, the United States Naval Observatory,
and the University of Washington.


\bsp

\label{lastpage}

\end{document}